\documentclass[AMA,STIX1COL]{WileyNJD-v2}
\pdfoutput=1

\usepackage{graphicx}
\usepackage{subcaption}
\usepackage{caption}
\usepackage{algpseudocode}
\algblockdefx[Case]{Case}{EndCase}[1][]{\textbf{case} #1 \textbf{of}}{\textbf{end case}}
\algloopdefx[CaseItem]{CaseItem}
[1][default]{\texttt{#1:} }
\usepackage{array}
\usepackage{url}
\usepackage{mypackage}
\usepackage{xcolor}
\usepackage{colortbl}
\usepackage{pgf}
\usepackage{tikz}
\usetikzlibrary{arrows,shapes,fit}

\newboolean{showcomments}
\setboolean{showcomments}{true} 

\ifthenelse{\boolean{showcomments}}
  {\newcommand{\note}[2]{
	\fbox{\bfseries\sffamily\scriptsize#1}
    {\sf$\Rightarrow$\textit{#2}$\Leftarrow$}
   }
  }
  {\renewcommand{\note}[2]{}
  }

\usepackage{todonotes}
\usepackage{paralist}

\graphicspath{{figs/}}

\articletype{Article Type}%
\received{}
\revised{}
\accepted{}
\begin{document}
\title{Caterpillar: A Business Process Execution Engine on the Ethereum Blockchain}

\author[1]{Orlenys L\'opez-Pintado}
\author[1,2]{Luciano Garc\'{\i}a-Ba\~nuelos}
\author[1]{Marlon Dumas}
\author[3,4]{Ingo Weber}
\author[3]{Alex Ponomarev}
\authormark{L\'opez-Pintado \textsc{et al}}

\address[1]{\orgdiv{Institute of Computer Science}, \orgname{University of Tartu}, \orgaddress{\state{Tartu}, \country{Estonia}}}
\address[2]{\orgdiv{School of Engineering \& Science}, \orgname{Tecnol\'ogico de Monterrey}, \orgaddress{\state{Puebla}, \country{Mexico}}}
\address[3]{\orgdiv{Data61}, \orgname{CSIRO}, \orgaddress{\state{Sydney}, \country{Australia}}}
\address[4]{\orgdiv{School of Computer Science \& Engineering}, \orgname{UNSW Australia}, \orgaddress{\state{Sydney}, \country{Australia}}}

\corres{ Institute of Computer Science,	University of Tartu, Juhan Liivi 2, 50409 Tartu, Estonia	     
	     \email{orlenyslp@ut.ee}}

\abstract[Summary]{
Blockchain platforms, such as Ethereum, allow a set of actors to maintain a ledger of transactions without relying on a central authority and to deploy programs, called smart contracts, that are executed whenever certain transactions occur. These features can be used as basic building blocks for executing collaborative business processes between mutually untrusting parties.
However, implementing business processes using the low-level primitives provided by blockchain platforms is cumbersome and error-prone. In contrast, established business process management systems, such as those based on the standard Business Process Model and Notation (BPMN), provide convenient abstractions for rapid development of process-oriented applications.
This article demonstrates how to combine the advantages of a business process management system with those of a blockchain platform. The article introduces a blockchain-based BPMN execution engine, named Caterpillar.
Like any BPMN execution engine, Caterpillar supports the creation of instances of a process model and allows users to monitor the state of process instances and to execute tasks thereof. The specificity of Caterpillar is that the state of each process instance is maintained on the (Ethereum) blockchain and the workflow routing is performed by smart contracts generated by a BPMN-to-Solidity compiler. The Caterpillar compiler supports a large array of BPMN constructs, including subprocesses, multi-instances activities and event handlers. The paper describes the architecture of Caterpillar, and the interfaces it provides to support the monitoring of process instances, the allocation and execution of work items, and the execution of service tasks.
}


\jnlcitation{\cname{%
\author{L\'opez-Pintado O.}, 
\author{B. Garc\'{\i}a-Ba\~nuelos}, 
\author{M. Dumas}, 
\author{I. Weber}, and 
\author{A. Ponomarev}} (\cyear{2018}), 
\ctitle{Caterpillar: A Business Process Execution Engine on the Ethereum Blockchain},
\cjournal{Software: Practice and Experience}, \cvol{2018;00:1--45}.}

\maketitle


\section{Introduction}
\label{sec:introduction}

Contemporary blockchain platforms allow a set of actors to maintain a ledger of transactions without relying on a central authority and to deploy programs, called \emph{smart contracts}, which can be called via \emph{transactions} by external actors and by other smart contracts to change the state of the ledger. For example, the Ethereum\footnote{https://www.ethereum.org/} blockchain network supports the execution of smart contracts coded in the Solidity language.

The combination of a tamper-proof decentralized ledger with smart contracts provides the basic building blocks to implement collaborative (inter-organizational) business processes involving mutually untrusting parties~\cite{WeberXRGPM16,DBLP:journals/tmis/MendlingWABCDDC18}. In fact, several existing blockchain applications implement business processes involving multiple independent participants, such as supply chain management processes~\cite{UKBlockchainReport,aureport}.
However, implementing business processes using the low-level primitives provided by blockchain platforms is cumbersome, error-prone, and requires specialized skills. According to a survey by Gartner~\cite{Gartner:2018:CIOSurveyBC}, around one fifth of relevant surveyed Chief Information Officers (CIOs) stated that one of the major roadblocks for the adoption of blockchain technology in their company is that it requires highly specialized teams that are the difficult to put together.

In contrast, established Business Process Management Systems (BPMS) provide convenient abstractions for rapid implementation of intra-organization business processes, by taking as starting point a business process model represented for example in the Business Process Model and Notation (BPMN)~\cite{bpmnspec} standard. These abstractions make it possible to implement and maintain process-oriented applications on the basis of process models, without requiring low-level or specialized development skills.

In this setting, the overarching question addressed in this article is the following: How to combine the high-level abstractions of BPMSs with the capabilities of blockchain technology, in order to support the execution of collaborative business processes between mutually untrusted parties?

Earlier work on blockchain-based business process execution, in part by the authors, provides some of the technical building blocks required to execute business processes on a blockchain platform. For example, Weber et al.~\cite{WeberXRGPM16,GarciaPDW17} described a compilation approach from a restricted subset of BPMN into Solidity smart contracts, while Prybila et al.~\cite{Prybila:FGCS:2017} discussed an approach to monitor business processes on a blockchain via specialized tokens. This article builds upon this earlier body of research and extends it into an open-source blockchain-based business process execution engine, named \emph{Caterpillar}. Like any typical process execution engine, Caterpillar supports the creation of instances of a BPMN process model and allows managers and process workers to track the state of process instances and to execute tasks thereof. The specificity of Caterpillar is that the execution state of each process instance is maintained on the Ethereum blockchain and the workflow routing is performed entirely by smart contracts generated by a BPMN-to-Solidity compiler covering the full spectrum of BPMN constructs, including subprocesses, multi-instance activities and event handlers, which are not supported by previous proposals. 

To the best of our knowledge, Caterpillar is the first prototype to demonstrate how a full-fledged business process execution engine can be deployed entirely on a blockchain platform in such a way that, once a process is deployed, no off-chain component is required in order to execute and monitor instances of the process.
Note that the goal of Caterpillar is not to replace intra-organizational BPMSs, but rather to provide the capabilities of a BPMS on blockchain. As such, it is intended to be used to implement inter-organizational processes, also known process-centric decentralized applications, or \emph{process-centric dapps} for short. Process-centric dapps are executed in environments characterized by lack of trust between parties, which entails a requirement to ensure that all parties comply with the rules captured in an agreed-upon process model. The goal of Caterpillar is to enable a set of parties to execute a process-centric dapp on a blockchain platform in a way that ensures \emph{compliance by design}, meaning that no party is able to execute a transaction that does not abide with the collaborative process model.

A preliminary version of the Caterpillar engine was summarily presented as a tool demonstration paper.~\cite{Lopez:BPM-Demo:2017} This article provides a detailed overview of the design principles, architecture, and implementation of this engine, including a detailed description of a mapping from BPMN to Solidity, which covers the full spectrum of BPMN constructs. The article additionally presents an evaluation of the Caterpillar engine, which demonstrates the tradeoffs between efficiency (specifically consumption of the cryptocurrency Ether) and the ability to run inter-linked business processes entirely on the blockchain -- without requiring external runtime components.

The remainder of this paper is structured as follows. 
Background and related work are discussed in \autoref{sec:bg-rel-work}, followed by a running example in \autoref{sec:example}. 
On this basis, we describe the design philosophy of Caterpillar and give and overview of the system and its architecture in \autoref{sec:architecture}.
\autoref{sect:compilation} delves into details of the compilation from BPMN to solidity.
Then we describe the implementation and evaluation in \autoref{sec:impl-eval} before \autoref{sec:outlook} concludes.


\section{Background and Related Work}
\label{sec:bg-rel-work}

In this section, we give relevant background on blockchain technology, and discuss related work.

\subsection{Background}
\label{ssec:background}

A blockchain is a distributed append-only store of transactions distributed across computational nodes and structured as a linked list of blocks~\cite{2019-Blockchain-Book}. Each block contains an ordered set of transactions.
A blockchain network is made up of nodes, a subset of which (called full nodes) holds a replica of the data structure, and in a public permissionless blockchain network the nodes can join and leave as they please.
Clients use a blockchain system (a concrete network) by reading data from and submitting transactions to it. Submitted transactions are grouped into blocks, which are broadcast across the network to be appended to the blockchain.
To be accepted, a transaction must be properly formed and signed by their creator. 
No trust in individual clients or nodes is required, as the transactions are cryptographically signed and validated, and broadcast widely.
A consensus mechanism ensures tamper-proofness without assuming mutual trust between participants.

Most public blockchains use \emph{proof-of-work} as consensus mechanism, whereby the creation of a new block, which is referred to as \emph{mining}, requires solving a computationally hard cryptographic puzzle. A node can choose to use its computational power for mining blocks. 
Mining the next valid block requires solving the puzzle before any other miner, and is rewarded financially.
Each block contains the hash value of the previous block, thus linking the blocks in the database. 
Moreover, any attempt to alter a block in the database would incur in high computational costs: to preserve the links established by the cryptographic hashes would require the recreation of the whole set of subsequent blocks from the altered block. 
The economic factor of block rewards and the resulting computational difficulty in altering a block makes tampering virtually impossible.

One of the novelties the Ethereum blockchain introduced are smart contracts.
A \emph{smart contract} is a computer program deployed on the blockchain, which may be invoked via a transaction\cite{2019-Blockchain-Book}.
When a new block is mined and broadcast, each full node in the peer-to-peer network is required to execute locally the set of transactions, including the ones calling smart contracts, and perform the respective calculations to derive the current state after execution of the transactions included in that block.
Applications that are designed to provide their main functionality through smart
contracts are called \emph{dapps}\cite{2019-Blockchain-Book}.

Smart contracts are executed over the Ethereum Virtual Machine (EVM), which is bundled within each peer node. The EVM is a runtime component, which provides a stack-based computing platform with a small set of operations that is sufficient to support the definition of Turing-complete programming languages. The size of the EVM word is of 256-bits. For each contract, the EVM allocates a persistent memory, referred to as storage, which is organized as a key-value store that maps 256-bit words to 256-bit words. The persistent memory is private such that it cannot be directly accessed by another contract or transaction. 
Moreover, a contract gets access to volatile memory with each function call, which can be expanded by one word at a time and that serves to store intermediate values. The EVM uses a stack and no registers to execute the instructions of the smart contract. The stack has a limit of 1024 words, and only the topmost 16 words are accessible at given moment in the execution, hence the need for volatile memory. Finally, each contract can write data into a log which is visible to external applications.

Several contract-oriented programming languages and compilers thereof have been developed that produce bytecode for the EVM. Among them, we have selected Solidity for our developments, because it is the most widely used and supported one. Solidity is a strongly-typed language and its syntax resembles the one of JavaScript. A contract in Solidity is defined in a similar way as classes in Java-like object-oriented languages. Thus, the definition of contracts usually includes persistent properties (i.e. the contract's state) and functions to query and manipulate the properties.

Each peer in the network communicates with other peers using the Ethereum wire protocol~\cite{wireprotocol}. The details of such protocol is out of the scope of this paper. In addition to the internal interactions, each peer exposes a number of methods over an RPC-style endpoint, which is known as the Ethereum's JSON-RPC API, because it uses data exchanges formatted according to JSON-RPC specification~\cite{jsonrpc}. It is this RPC endpoint that external applications (e.g. wallets and other software) use for interacting with the Ethereum blockchain.

From a technical point of view, a (smart) contract corresponds to code that is deployed to the Ethereum blockchain. Once deployed, the contract is associated with an address and a working memory. By knowing a contract's address, any external application can execute the public functions defined by a contract. Generally speaking, the interaction between an external application and a contract can happen in two ways. On the one hand, an external application requests the execution of a transaction by calling a contract's function. Such transaction is then forwarded to the network of peers. A transaction is seen as executed only if a block includes the transaction in the database. Since this type of interaction requires block mining, on public blockchains the requester is typically required to pay a transaction fee. On the other hand, an external application can request the execution of a contract's function on a single Ethereum node, that is, without forwarding a transaction to the network of peers. Since no block mining is required, this type of interaction incurs no fee. This type of interaction can be used for querying the current contract's working memory state or for previewing the outcome of executing a contract's function given the current state. Moreover, external applications cannot access to the contract's working memory state unless the contract provides public functions permitting it. 

In contrast, smart contracts have no way to call external programs. However, as mentioned before, a contract can write information in a log that is visible to external applications. Moreover, the JSON-RPC API provides some operations that can be used to install log-filters on a local peer that can be repeatedly polled to retrieve the entries added to the contract's log. This way of interaction is widely used to implement a sort of push-oriented interaction with external applications and to forward requests to so-called blockchain oracles, as discussed later on.

\subsection{Related Work}
\label{ssec:rel-work}

Weber et al.~\cite{WeberXRGPM16} propose an approach for model-driven engineering (MDE) of collaborative business processes on top of the Ethereum blockchain platform. In this proposal, the starting point for implementing a business process is a BPMN\cite{omg2013bpmn} choreography diagram. A choreography diagram is a model of a collaborative process consisting 
of a number of message exchange tasks (called choreography tasks) and control-flow routing constructs (specifically XOR and AND gateways). Weber et al.\ propose to include additional tasks for payments and data transformation, in extension of the BPMN standard.
The parties in a choreography interact via message exchanges, which are sent as transactions on the blockchain in the approach by Weber et al. The authors propose to compile a BPMN choreography diagram into Solidity contracts that ensure that the parties can only record their message exchanges in a way that is compatible with the ordering relations captured in the BPMN choreography diagram. 

Prybila et al.~\cite{Prybila:FGCS:2017} present an alternative approach to monitor business processes executed on top of the Bitcoin blockchain via specialized tokens. Like in Weber et al.~\cite{WeberXRGPM16}, the authors assume that the collaborative process is modeled as a choreography. In other words, both of these approaches assume that the parties in a collaborative business process interact via message exchanges, and they use the blockchain platform to record the message exchanges and to check or enforce that these exchanges occur in a certain order. 
This effectively means that the blockchain platform serves as one execution component of a collaborative process, but many of the components are off-chain, and the interactions between parties occur via message exchanges. 


In some of our prior work~\cite{GarciaPDW17}, we propose an approach to transform BPMN process diagrams into Solidity smart contracts. This latter work does not assume that the parties communicate via message exchanges -- but instead the parties use the blockchain party as a coordination mechanism to maintain the state of the process and to determine what tasks or events may occur next given the current state. The emphasis of this latter work is on optimizing the generated Solidity code in order to reduce the costs related to deployment and execution of smart contracts. However, this work is restricted to flat BPMN process models (no subprocesses) consisting only of tasks and basic gateways (AND and XOR gateways). This previous approach does not support subprocesses, boundary events, or multi-instances activities.

Hull et al.~\cite{Hull:ICSOC:2016} discuss a vision of how business process modeling, and more specifically the Artifact-centric paradigm~\cite{Nigam:IBM-SJ:2003} would be a suitable approach to model collaborative business processes executed on top of blockchain technology. Similarly, Norta~\cite{Norta2015} advocates the use of blockchain to coordinate collaborative business processes based on so-called choreography models, which are similar to collaboration diagrams in that they rely on the assumption that parties interact via message exchanges. Another related work~\cite{Frantz:ECAS:2016} proposes a mapping from a domain specific language for ``institutions'' to Solidity. These previous research efforts only outline a possible architecture for modeling and executing blockchain-based business processes, but they do not provide any implementation or evaluation.
In contrast, Sturm et al.~\cite{sturm2018lean} provide an implementation, but their approach expresses control flow solely via ``requires'' relationships, i.e., specifying which tasks are required to be completed before a given task can be executed. 
This allows expressing AND and OR gateways, but does not disable alternative, un-executed tasks when executing an OR join\footnote{\url{https://github.com/Jonasmpi/PExSCo/blob/master/contracts/ContractCollaborationManager.sol}, commit a26274c9d564f8a2cc2477eaa26f633b485323ca, lines 107-122; accessed 3/12/2018}.
By comparison, Caterpillar follows a different approach and supports a much wider range of BPMN elements, including subprocesses, multi-instance activities, boundary events, and several types of BPMN tasks.

Lorikeet\cite{2018-Tran-BPM-Demo} is an MDE tool that implements the approach of Weber et al.\cite{WeberXRGPM16} and the translation algorithm from Garcia et al.\cite{GarciaPDW17}.
The process components are supplemented with asset registries (e.g., tokens representing coins or titles). 
Lorikeet has been successfully applied in a number of projects with industry\cite{2018-Tran-BPM-Demo}, demonstrating the value of process-oriented smart contract generation.
The emphasis of Lorikeet is on the MDE approach, where the generated code can be used as a basis for the implementation; Lorikeet is not a BPMS.

The idea of executing business processes using blockchain and smart contract technology has also been considered by industry practitioners and BPMS vendors. For example, Rikken~\cite{Rikken:bpmleader.com:2015} discusses some perceived advantages of executing business processes via smart contracts. Bonitasoft~\cite{bonitasoft} provides a software connector that enables process instances running on the Bonita BPMS to execute transactions on a blockchain. Using this connector, a task in a BPMN process model can be configured in such a way that its execution generates a transaction on a blockchain. 
Similarly, \cite{IBMBlockchainBPM} shows how to use IBM's BPM system to execute business processes on top of the Hyperledger blockchain platform. In this latter approach, data objects are kept in a (permissioned) blockchain, and tasks in the process read and write into these objects.
This approach is suitable when one or a handful of tasks in a process need to execute transactions involving untrusted parties, or need to leave a tamper-proof trace of their execution. 
When permissioned blockchains are used (which are more scalable than public ones), it becomes practical to use this approach to record every execution event (e.g.\ events indicating the start and end of each task) on a blockchain. However, in this approach, the blockchain stores the execution trace of tasks in a process, and possibly also the data produced by these tasks, but it does not ensure that the execution of the collaborative process abides to a given process model. The process is executed within the (off-chain) BPMSs of the actors involved in the process, and these systems may perform tasks even if the current state of the collaborative process does not allow so.
Hence, this approach is not suitable when the goal is to implement an end-to-end collaborative business process in a way that benefits from the integrity properties of blockchain technology.


In summary, we observe that, except for our previous work~\cite{GarciaPDW17}, existing approaches for executing collaborative business processes on top of blockchain technology use the blockchain to record message exchanges or transactions and use smart contracts to check and/or enforce that messages are exchanged in a way that is compatible with a collaborative process model. In other words, most parts of a business process are still executed in the BPMSs or other process-aware information systems of each business party, and the blockchain is used to record, monitor, and occasionally control interactions between the processes executed by each party.
Also, existing approaches focus on a highly restricted subset of BPMN, comprising tasks, events, and XOR and AND gateways. In particular, these approaches do not support the execution of hierarchical process models (i.e. processes linked with subprocesses).
Therefore, this article advances the state of the art in the field with a proposal of a process execution engine, Caterpillar, with broad support of BPMN elements. 
Our approach enables the implementation and execution of collaborative business processes, where all critical components of the BPMS are hosted on the blockchain.


\section{Running Example}
\label{sec:example}

\begin{figure}[htp]
	\centering
	\begin{subfigure}[b]{\textwidth}
		\centering
		\includegraphics[scale=0.45]{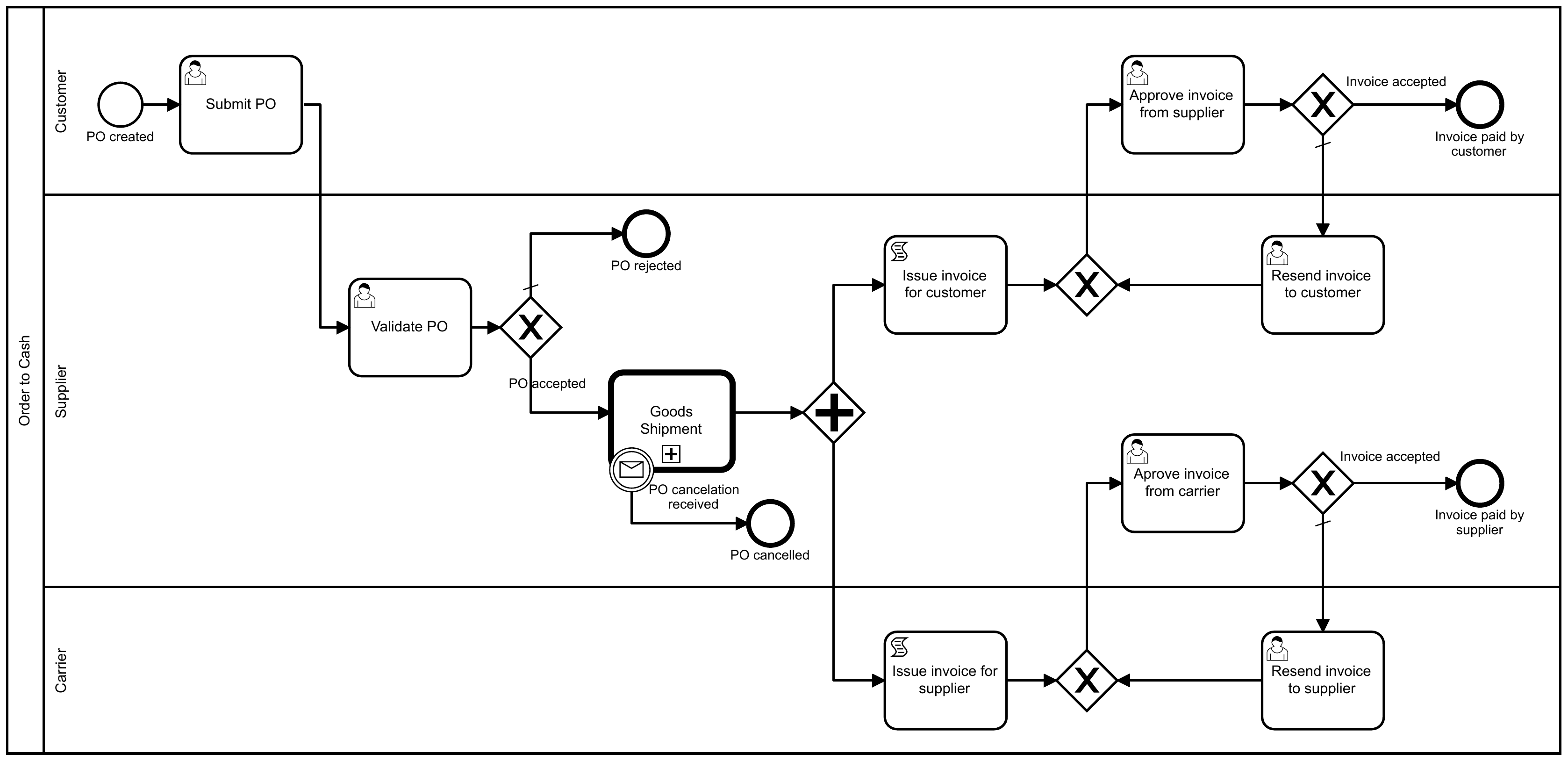}
		\caption{\label{subfig:sub1}}
	\end{subfigure}
	
	\begin{subfigure}[b]{\textwidth}
		\centering
		\includegraphics[scale=0.45]{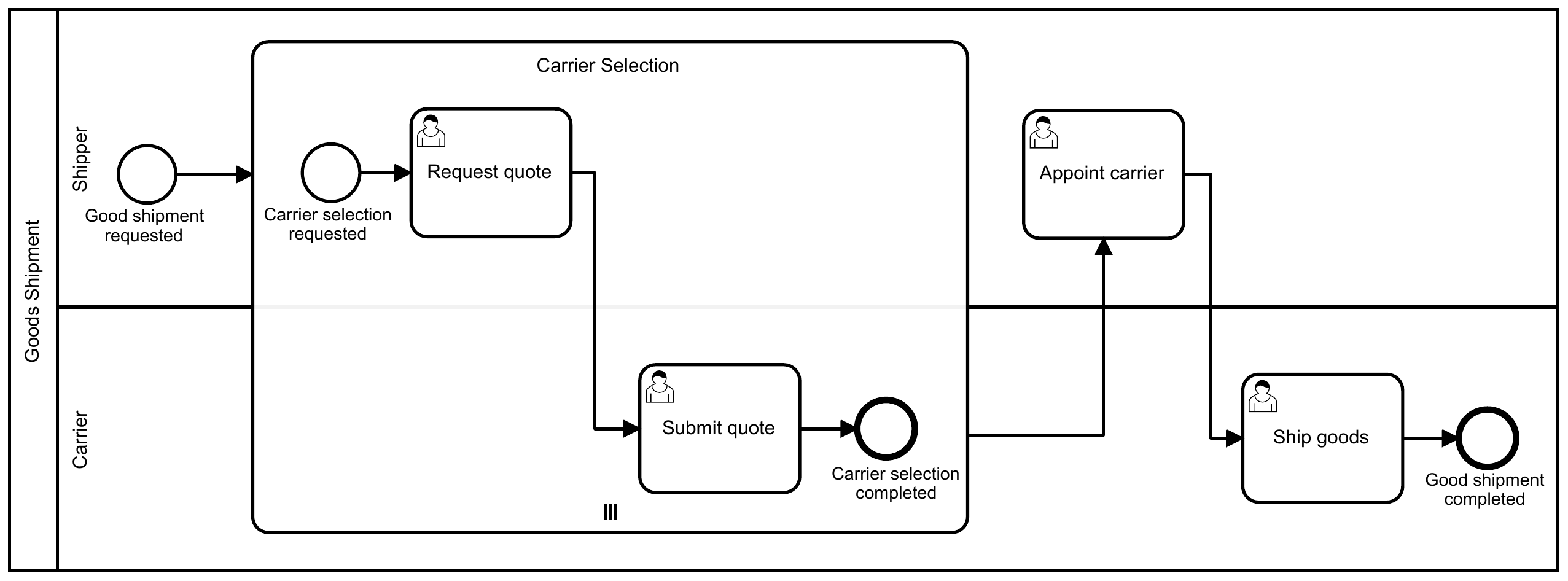}
		\caption{\label{subfig:sub2}}
	\end{subfigure}
	\caption{\label{fig:running}Running example: An order-to-cash process 
		(\subref{subfig:sub1}), with a shipment subprocess (\subref{subfig:sub2})}
\end{figure}

In the following, we introduce a running example that will be used throughout the paper.
\autoref{fig:running} presents the business process to illustrate concepts and Caterpillar's components. For convenience, the business process is modeled into two separate diagrams: an {\sc Order to Cash} root process model and a reusable {\sc Goods shipment} subprocess, which is called from the root process model. 

%

The process starts with a plain start event. This is a requirement as Caterpillar only supports so-called \emph{explicit instantiation}, i.e.\ it does not support implicit instantiation via start timer events or start message events.

The process execution starts with the user task {\tt Submit PO}. A user task is specified with rounded rectangle decorated with person like icon. A user task adds a workitem into a stakeholder's work list. 
The user task {\tt Submit PO} is intended to let a stakeholder enter the details of the purchase order.
Next, the case continues with user task {\tt Validate PO}. The intention of this task is to allow a stakeholder to check the validity of the PO, the existence of the goods
in the warehouse and to submit a decision over the PO (e.g. PO is accepted or rejected).
Next, the execution of the process proceeds reaching an exclusive gateway (e.g. diamond 
decorated with a $\times$), which selects one out of the two possible execution paths
based on a predicate that is formulated over the process data, i.e. the decision taken
by the stakeholder in regards of the validity of the PO.
When the PO is rejected the execution flow will reach the end event {\tt PO reject}, ending 
the execution of the process instance.
On the other hand, if the PO has is accepted the execution flow will reach the call 
activity {\tt Goods Shipment}. Executing a call activity implies the instantiation of
the {\sc Goods Shipment} subprocess.
The boundary event attached to the call activity {\sc Goods Shipment} indicates that the customer has the right to cancel 
the order so long as the goods have not yet been dispatched.

The execution of the subprocess {\sc Goods shipment} eventually reaches a point where
multiple instances of the subprocess {\sc Carrier selection} are created. The intuition behind
is that each instance of the subprocess implements the interaction between supplier
and carrier companies to get a quote\footnote{The demonstration process included in 
Caterpillar's code repository creates a fixed number instances of the {\sc Carrier selection} 
subprocess. The number is specified with a process variable and is set to two by default.}.
When all the quotes are submitted by the carrier companies, 
a clerk on the supplier company selects one carrier based on the quotes and organizes 
the shipment. When the user task {\sc Ship goods} finishes, the execution flow is passed
back to the root process. 
The process instance proceeds by activating the two parallel paths to process the 
payment of the shipment by the supplier and the payment by the customer.
At this point, the process includes two script tasks, rounded rectangles with a folded paper-like
icon, which represent Solidity scripts to issue invoices.
Once the invoices are issued, it is expected that the customer and supplier pay their
corresponding invoices.
Note that 
the process considers the possibility of reissuing the invoices, e.g. when one invoice
is wrong.
The overall process ends when the parallel paths reach their corresponding end events.


\section{Design Principles and System Overview}
\label{sec:architecture}
\subsection{Design Principles}
\label{ssec:designphilosophy}

In the context of collaborative business processes, blockchain technology allows us to ensure that the parties involved in a collaboration comply with an agreed-upon collaborative process model. For example, in a collaborative process involving a purchasing company, a supplier and a carrier, blockchain technology allows us to ensure that the carrier does not submit the invoice to the supplier before the delivery has been acknowledged by the purchasing company. 

Compliance with respect to a (collaborative) process model can be ensured using one of two approaches~\cite{SadiqGN07}. The first approach is \emph{compliance by monitoring}, which means that the parties record their transactions on a blockchain so that all other parties can check that process has been executed as agreed. When a party deviates with respect to the agreed-upon process, other parties can detect it and trigger a conflict resolution procedure. This approach has been taken by several commercial Business Process Management Systems (BPMSs) such as Bizagi and Camunda, which offer adapters to record the transactions produced by a process in a blockchain. The second approach is \emph{compliance by design}. In this latter approach, the parties execute each step of the business process by invoking a transaction on the blockchain. When a transaction is invoked, the blockchain platform checks the current state of the process and the inputs and outputs of the transaction. The transaction is accepted if and only if it complies with the collaborative process model. This approach requires that the full specification of the collaborative process is encoded as smart contracts running on the blockchain. Caterpillar is an embodiment of this compliance by design approach.

The compliance-by-design approach is suitable when the level of trust between parties is low, the impact of non-compliance is high and the cost of conflict resolution is high (e.g. if the parties are in multiple jurisdictions). This is the scenario addressed by Caterpillar. Conversely, when the parties have some minimum of level of trust, the impact of non-compliance is limited, and conflict resolution is straightforward, an approach based on compliance by monitoring is more suitable and hence a BPMS that uses the blockchain purely as a secure logging mechanism is sufficient. 


To recap, the goal of Caterpillar is to enable a set of parties to develop, deploy, and execute a process-centric dapp on a blockchain platform in a way that ensures \emph{compliance by design}, meaning that no party is able to execute a transaction that does not abide with the collaborative process model. Naturally, Caterpillar also seeks to make the development and deployment of the process-centric dapp as seamless as possible, by starting from a high-level specification of the collaborative process and automating the compilation and deployment of this specification into the blockchain platform. 


In line with the above, the design of Caterpillar is driven by the following principles:
\begin{enumerate}
\item 
The collaborative process is modeled in the same way as an intra-organizational business process executed on top of a traditional BPMS. In other words, the collaborative process is modeled as if all the parties shared the same process execution infrastructure (the blockchain). Accordingly, the starting point for implementing a collaborative business process is a single-pool BPMN process model (not a collaborative process or a choreography where parties communicate via messages). Each independent party in the process is represented as a lane. Hand-offs between parties are simply represented via sequence flows that go from one lane to another (and not via messages). 
\item A collaborative process model may comprise subprocesses. Accordingly, an instance of a process may be linked to instances of subprocesses and vice-versa.
\item 
The full state of the process instance is recorded on the blockchain, and all the metadata required to retrieve the links between a given process instance and its related subprocess instances is also recorded on the blockchain.
\item
All the execution logic captured in the process model is translated into smart contract functions, which can run independently of any other runtime component. In other words, the execution of process instances can proceed even if no instance of the off-chain runtime component is running. Also, several instances of the runtime component can be running at a given point in time (e.g. one instance per participant).
\end{enumerate}

The rationale behind the first point above deserves some further explanation.
First, in the blockchain setting BPMN collaboration diagrams could be used, with message exchanges between parties (represented as pools). That would be suitable if blockchain was primarily used in its function as data store (for messages) and communication mechanism (transaction and block broadcast) -- but would keep the process execution logic and business rules off-chain, which may incur on security/trust issues as they can be tampered. 
Alternatively, a smart contract on the blockchain could be represented as another pool, which would be the orchestrator. In that case, all messages from participants would be relayed by the smart contract, possibly with computation by it (such as script tasks). However, that notation would be clumsy, and more of a pure implementation artifact rather than a business process model that serves as a notational bridge between business experts and developers.
Another alternative would be BPMN choreography diagrams. However, this standard was devised with the premise that ``there is no central controller, responsible entity, or observer''\cite[p. 23]{omg2013bpmn}, which exactly opposes the role smart contracts and blockchain can play.
For this reason, we decided to use process diagrams with a single pool for Caterpillar: It can represent the business process, and makes use of blockchain for data storage, computation, and communication.


\subsection{Architecture Overview}

\begin{figure}[bt]
	\centering
	\includegraphics[width=0.8\textwidth]{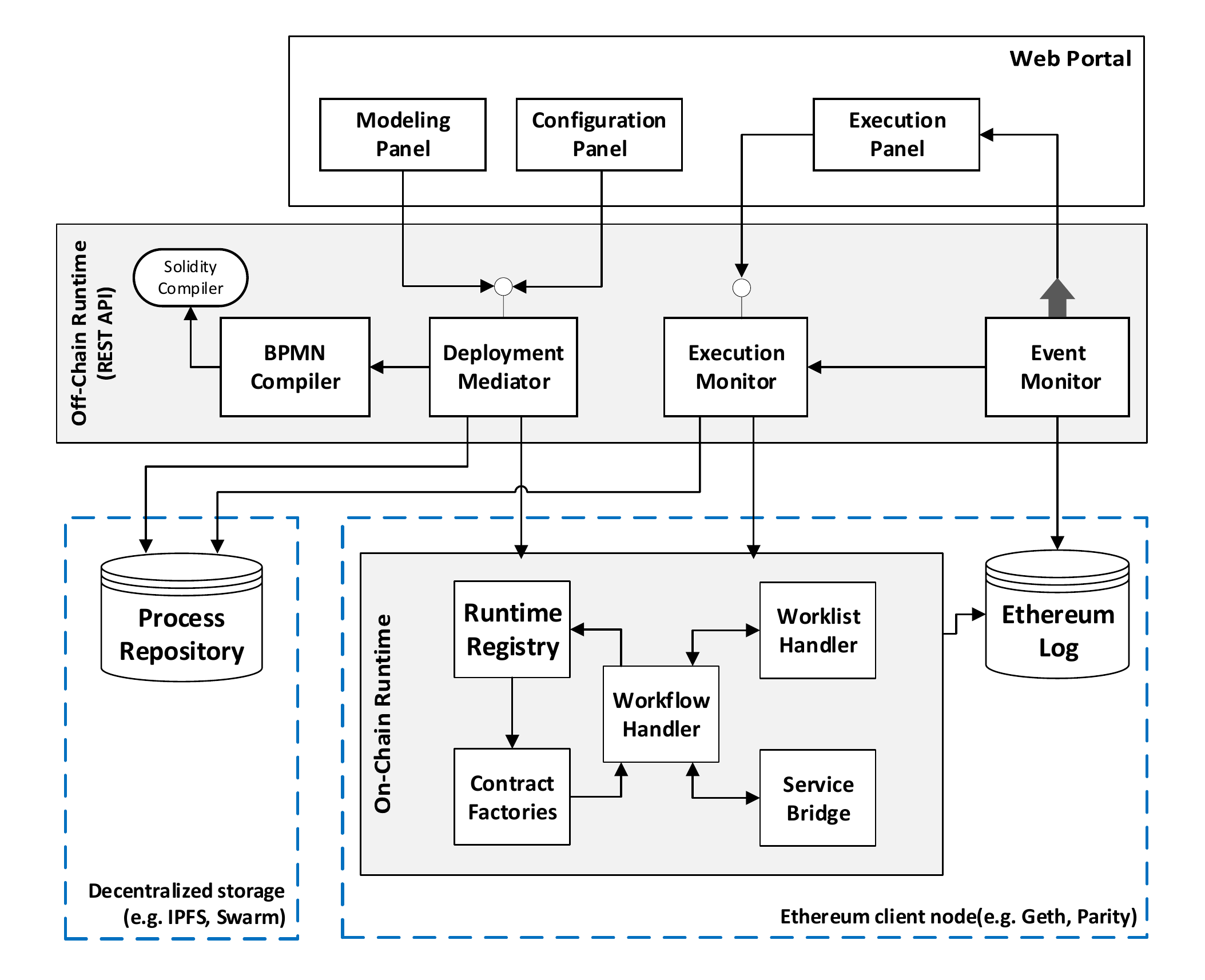}
	\caption{\label{fig:architecture}Architecture of Caterpillar}
\end{figure}

The above principles guided the design of Caterpillar's architecture, which is organized into three layers as shown in \autoref{fig:architecture}.
The layer at the bottom will be referred to as the ``On-chain Runtime and Storage'' layer.
Specifically, the ``On-chain Runtime'' refers to a set of smart contracts that includes
housekeeping support code, e.g. process instantiation, as well as 
process specific code, e.g. control flow, process data, etc.
The on-chain parts are replicated across all full nodes of the blockchain network, e.g., the public Ethereum network.
``Storage'' refers to the ``Ethereum log'', which serves to externalize process events and data, and the ``Process repository'', which keeps compilation artifacts and the like.

In the middle, the ``Off-chain Runtime'' layer includes a set of tools to compile, deploy and monitor business processes on the On-chain layer. Specifically, the Off-chain runtime includes a BPMN compiler, a deployment mediator, an execution monitor and an event monitor. Note that these Off-chain runtime components can be hosted by each actor in a collaborative process separately. In particular, it is not necessary for all actors to use the same Event Monitor or Execution Monitor. Following the design principles introduced above, the state of all process instances is stored on-chain and all the decision points are evaluated on-chain. Hence, if one of the event monitors tries to execute an event at the wrong moment, it would be rejected by the corresponding smart contract. Similarly, each actor can host its own BPMN Compiler to generate the smart contracts implementing a process model -- and in this way they can cross-check the contacts deployed on the chain. Also, each actor can use their own deployment mediator, though on the end, only one of them should deploy a given process (the others can check the deployed contracts).


%
Finally, the top-most layer comprises a set of components for editing executable process
models, packaging process configurations (e.g. adapting control flow code with components
implementing different resource management schemes, etc.), and to monitor the execution
of process instances/cases.

Below, we provide a more detailed description of the components in each one of these layers.

\subsection{On-Chain Runtime and Storage}
 
The dashed rectangles in \autoref{fig:architecture} divide the bottom layer of the architecture into two parts. On the right, the ``On-chain Runtime'' components running on the Ethereum blockchain platform store and support the execution of smart contracts that fully encode a set of process models. The events generated by these smart contracts are recorded in the blockchain platform's log, which is accessible from outside the blockchain. 
On the left, a ``Process Repository'' is used to keep data received, produced, and required by Caterpillar to execute the process models deployed on the blockchain. Below we discuss these two parts in turn.

\subsubsection{Process Repository and Ethereum Log}
 

The ``Process Repository'' (bottom, left-hand side of \autoref{fig:architecture}) stores and provides access to compilation artifacts, including the BPMN process models, the Solidity code generated therefrom, and addition metadata to link the generated Solidity code to elements of the BPMN models. This metadata is used during the deployment of Solidity smart contracts (when a new process instance is created) and also to link the  state of a running process instance to the corresponding BPMN model.
The Process Repository is implemented on top of the InterPlanetary File System (IPFS)\footnote{https://ipfs.io/}.
We note that the compilation artifacts in the process repository could be stored directly on the blockchain as a stream of bytes inside a smart contract. However, this alternative approach would entail a high storage cost, given that the storage is immutable for the lifetime of the blockchain. IPFS provides an alternative decentralized approach to storing the compilation artifacts at a lower cost, while producing an immutable and unique cryptographic hash key that uniquely identifies each of the compilation artifacts, and ensures absence of manipulation (albeit without guaranteeing availability).


The process-related smart contracts generated by Caterpillar store the following data on-chain for each process instance: (i) the state of the process instance from a control-flow perspective, in order to determine which tasks are enabled/started; (ii) the data that needs to be given as input to the tasks of the process; and (iii) the data required to evaluate the conditions in the decision gateways. Beyond these minimum requirements, a developer may specify additional variables in the BPMN model, in which case these variables are also stored on-chain, but this is optional. As an alternative and in order to strike a tradeoff between costs and tamper-proofness, it is possible to store a link to the full data objects on-chain (and possibly also a hash code for verification) and keep the full data off-chain, or to keep the full data objects in IPFS, for example by appending it to the Process Repository. This latter approach can also be applied when the volume of data that needs to be given as input to user tasks and service tasks is too large. In other words, the full data objects required by these tasks can be stored off-chain or in IPFS, so that only a link and a hash code need to be stored on-chain. A discussion on the performance, availability, and security properties of these alternative design decisions for data storage can be found in~\cite{2019-Blockchain-Book}.


The Ethereum log (right-hand side of \autoref{fig:architecture}) provides a medium for communication between off-chain and on-chain components. Specifically, when a transaction is included in the Blockchain, an event can be emitted and then written in the log to notify external components that some updates took place, e.g., to announce that some task became enabled and a process participant can execute it. This mechanism is used to manage interactions between the process-related smart contracts generated by Caterpillar, and external resources such as software services, as discussed in \autoref{sect:compilation}.

\subsubsection{On-Chain Runtime Components}

The ``On-chain Runtime'' consists of five components as shown in \autoref{fig:architecture}.
First, the ``Workflow Handler'' comprises the set of smart contracts generated by Caterpillar from the input BPMN models to handle the control-flow of the process models. The next  two components named ``Worklist Handler'' and ``Services Bridge'' 
consist of smart contracts 
that enable the interaction with external applications and to validate any data checked-in into
to the process instances, thus managing the interactions with external applications and users as
specified by user tasks and service tasks in the BPMN model.
The Worklist handler is responsible for managing user tasks (i.e. tasks to be performed by end users in BPMN) while the Service Bridge handles service tasks in BPMN, i.e. programmatic interactions with external applications exposed as services.
The Worklist Handler and Service Bridges are  implemented similarly: they consist of a smart contract that acts as mediator for forwarding a request  (via solidity events) and receiving the corresponding response (via a contract function call). Caterpillar provides a simple and generic implementation of the Worklist Handler, which keeps track of work items that are enabled, started, or completed. 
In contrast, the ``Service Bridge'' is a set of smart contracts that must be provided together with the BPMN process model, because Caterpillar cannot generate it as the details of these contracts are dependent on the services to which the bridging is made. 
The bridge contracts need to implement a program interface specified by Caterpillar.

The component ``Contract Factory'' includes a set of contracts that serve to
instantiate the smart contracts associated with a BPMN model as required: it instantiates
the smart contracts implementing the ``Workflow handler'' and ``Worklist handler'' of the root process, binds them and then
fires the execution of the process instance.

\begin{figure}[hbtp]
 	\centering
 	\includegraphics[width=\textwidth]{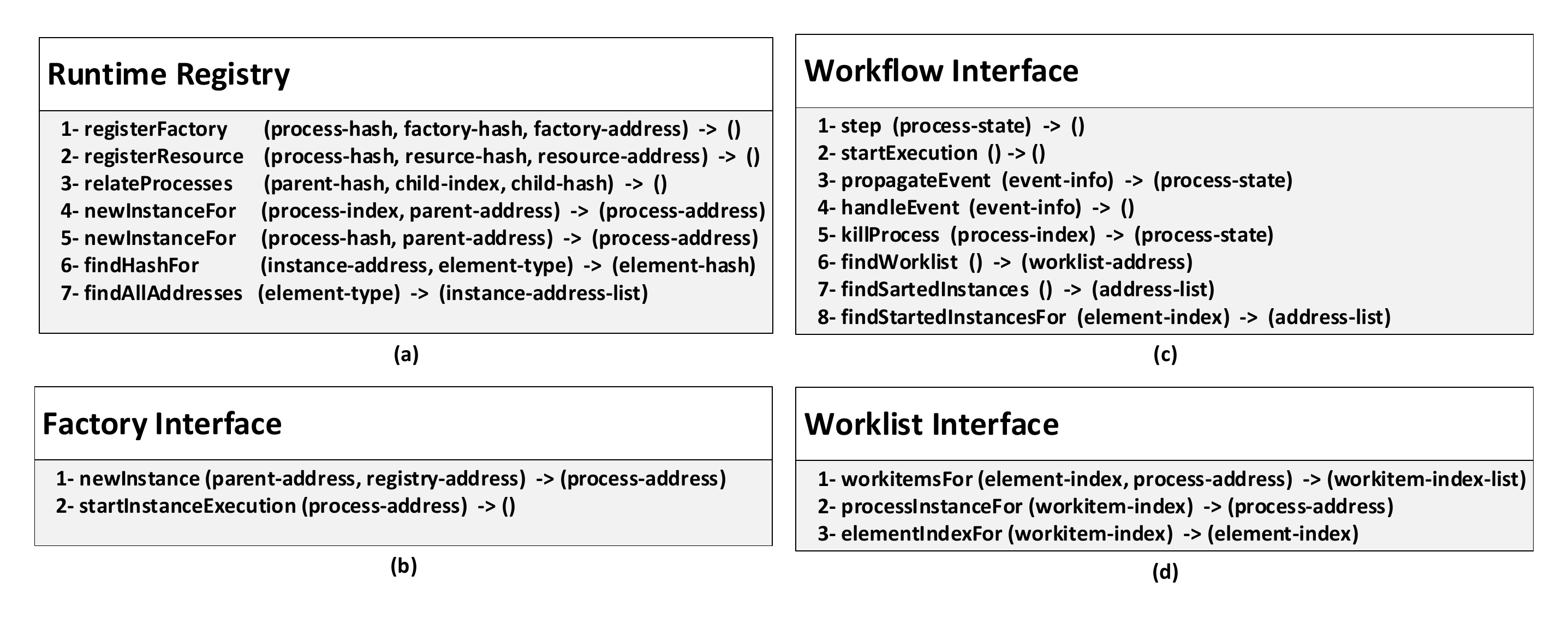}
 	\caption{\label{fig:on-chain-classes} Interfaces with their operations in the smart contracts managed by Caterpillar. }
 \end{figure}

The fifth on-chain runtime component, namely the ``Runtime Registry'', is a smart contract that keeps track of process 
instances (addresses of deployed ``Workflow handler'' smart contracts) and 
their relation with other smart contracts in Caterpillar's on-chain runtime.
The main functionalities of the ``Runtime Registry'' 
are shown in \autoref{fig:on-chain-classes}(a).
The suffixes on the name of the parameters hint at the location of the underlying
artifact: when the artifact refers to metadata stored in IPFS, then a hash is used
as identifier; when the artifact refers to smart contracts, then the identifier
corresponds to the address of the deployed contract; and when parameter name
uses an index as identifier, then a level of indirection (e.g. a solidity mapping)
is used.


 
\autoref{fig:on-chain-classes}(b)-(d) also introduce the interfaces implemented by the three typical contracts produced by the ``BPMN Compiler''. The functions in \textit{Workflow Interface} can start the execution of an instance, perform internal steps updating the process status according to updates, throwing and catching events, finish running instances, as well as finding information of related resources and contracts, e.g. invoked through call activities. 
On the other hand, the \textit{Worklist Interface} provides functions to access the information of workitems. A workitem, identified by an integer index, includes the address of the related control flow contract and the index of the activity to be performed by a process participant. Also, contracts implementing both interfaces above include a set of functions relying on the elements extracted from process models (see \autoref{sect:compilation} for further details). 
Finally, factories must implement two methods. The first function instantiates the corresponding process, taking as input the addresses of the ``Runtime registry'' and the contract which is responsible for creating the new instance (zero if an off-chain component executes the operation). The second function forces that process executions must be started explicitly, not when creating the new instance in the blockchain. Caterpillar automatically executes some elements internally at the moment they are enabled as explained in \autoref{sect:compilation}. Therefore, any contract instantiated from an element, e.g. a call activity, must be deployed before starting the execution when the respective element has been reached.  
 
 Before creating new instances of a process, the corresponding factory contract 
 must be deployed on Ethereum. 
 The address of the process factory must be stored/associated with an identifier to a process model in the repository. 
 The factory of this model-factory mapping can be updated at any time using the function {\tt registerFactory} which allows a process model to be implemented using different strategies. Besides, as interfaces 
 offer no information about the concrete implementation of the factory, an identifier 
 to recover such information from the repository must be provided. A similar approach 
 applies to linking a process with the contracts supporting the interaction of 
 participants, i.e. worklist for humans or services for information systems, using the 
 function {\tt registerResource} in both cases. Relations of processes with call 
 activities and other BPMN elements linked/mapped into smart contracts are also stored, 
 using the function {\tt relateProcess}. In this case, parent refers to the process 
 containing the element with the link. Unlike previous relations, here the parent will 
 instantiate a child given its index, but beforehand the referred contract 
 must be deployed. Thus the corresponding identifier is required for the registration.

 Caterpillar forces the process instantiation to be made through the ``Runtime Registry''. 
 To that end, two alternatives are available as functions {\tt newInstanceFor}. 
 An off-chain component must provide the identifier of the process as shown for the first 
 variant of the function. In this case, the parent address is zero because Caterpillar 
 prevents external actors from creating process instances linked to other process instances -- this mechanism is internal to Caterpillar. 
 In other words, external actors only can instantiate root processes, never subprocesses directly. 
 Indeed, a parent process has to create instances of a child given the integer index of 
 the BPMN element linked to the corresponding contract. In both cases, the 
 ``Runtime Registry'' verifies if the required contracts were deployed and instantiated 
 (e.g. factory, worklist, service), and then chooses the factory accordingly. Once the 
 instance is created, its address is published in the event log of Ethereum to notify 
 external actors that the subprocess instance has been created and how to reach it. 
 Besides, the ``Runtime Registry'' keeps a record of this new address that is associated 
 with the process identifier.
 
 The ``Runtime registry'' also provides methods to retrieve the identifier of a process running at a given address. Note that this identifier serves to find and check the information related to the process and its compilation metadata from the repository. Besides, the list of all the addresses created for any category of contracts can be recovered from the registry. In \autoref{fig:on-chain-classes}(a) the parameter \textit{element-type} refers generically to the different categories of contracts, e.g. factories, worklist, service and process models. 
 In summary, the ``Runtime registry''\footnote{A full implementation of the ``Runtime Registry'' can be accessed from \url{https://github.com/orlenyslp/Caterpillar/tree/master/v2.0/caterpillar-core/src/models/abstract}.} provides full control of the processes deployed by Caterpillar; the history of process executions, active instances, relations between contracts, and the metadata required to create new instances can be retrieved.

\subsection{Off-chain Runtime}

The off-chain runtime component of Caterpillar provides a service- oriented layer that allows external applications to interact with the on-chain components and the repository. The off-chain components enable external applications to compile a process models into Solidity smart contracts, to deploy said smart contracts, to query the status of process 
instances, and to register the execution of tasks associated with active process instances. Accordingly, the off-chain runtime consists of three modules: a BPMN compiler (which uses a Solidity compiler), a deployment mediator, and an execution monitor. The latter component relies on a fourth component, namely the Event Monitor, which listens for relevant execution events from the Ethereum log.

The runtime off-chain components are optional, i.e.\ the parties can directly invoke the smart contract transactions without going through Caterpillar's off-chain components, or they can implement their own runtime. The Caterpillar on-chain components ensure that, should an off-chain component be tampered, this would not affect the integrity of the process execution recorded on the blockchain, since the transaction corresponding to a task is executed if and only if the current state of the corresponding process instance allows so.

However, the tampering of an off-chain component may affect what the parties observe (e.g. the notifications they receive). The involved parties can mitigate this vulnerability by introducing an additional secured software component that queries the blockchain to check that a task is enabled before executing it. For example, if a carrier receives a notification from the off-chain component indicating that the products should be shipped, the additional secured layer should check that the ``Ship products'' task is enabled via the on-chain components before performing this task. In this way, a potential tampering of the off-chain components would be detected, and the goods would not be shipped. 
 
\subsubsection{Compiler}

\begin{figure}[hbtp]
 	\centering
 	\includegraphics[scale=0.85]{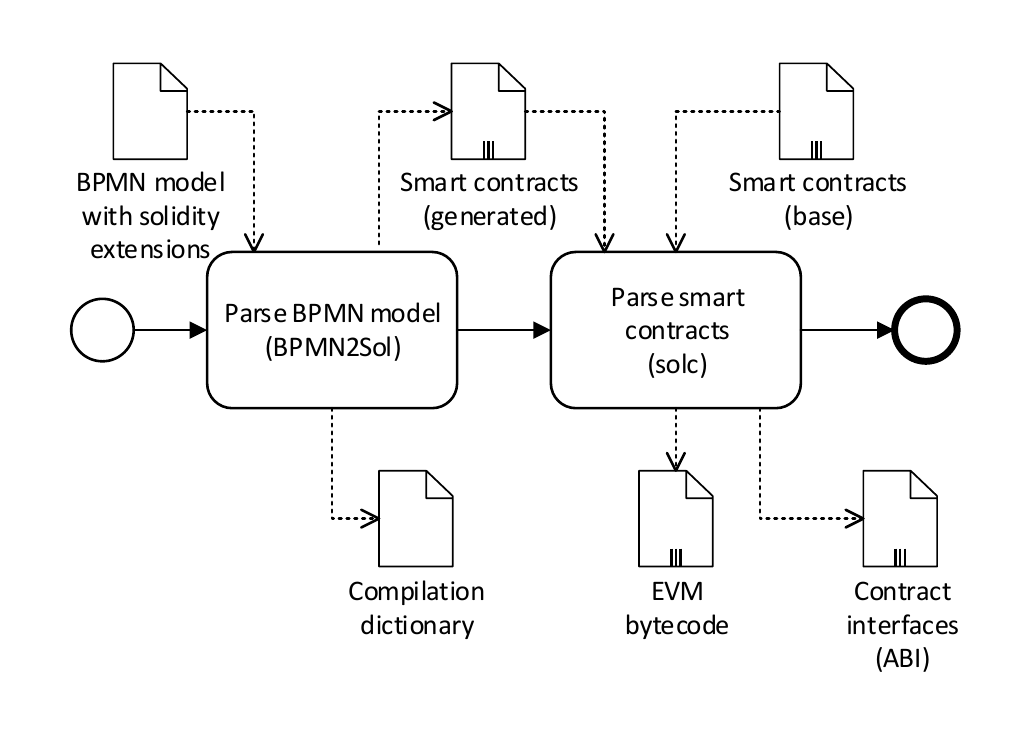}
 	\caption{\label{fig:compilation}Caterpillar's compilation process}
 \end{figure}

The first off-chain component, the BPMN compiler, is responsible for 
compiling BPMN process models into smart contracts.
This compilation is done in two steps (cf.~\autoref{fig:compilation}). First, 
the BPMN process model is compiled into a set of Solidity smart contracts 
plus additional metadata, called the \emph{compilation dictionary}, which is 
used later for monitoring purposes. The compilation dictionary is a data 
structure that includes information to map the elements in the BPMN model to 
the generated code. This information includes the name of contract method 
associated with any activity, a unique integer index assigned to each 
element, as well as the respective element type.

 
 
In the second step of the compilation process, Caterpillar puts together the 
set of smart contracts produced in the first step and, if applicable, the set 
of already existing contracts such as the ones corresponding to the 
interfaces of service tasks and call activities. These Solidity smart 
contracts are passed to the solidity compiler which produces EVM bytecode, 
and ABI\footnote{ABI (Application Binary Interface) is a 
JSON-based description of the list of the public methods implemented by a 
solidity smart contract as well as their signature. Several language bindings 
use the ABI definition, e.g. web3js in the case of JavaScript, to enable the 
interaction with smart contracts deployed into Ethereum.} definitions that 
are used for deploying the smart contracts to Ethereum. The ABI definitions 
are further used by Caterpillar's off-chain runtime or any other third-party 
application to interact with the deployed contracts. Artifacts involved in 
the compilation process, that is, input BPMN models, solidity contracts, 
compilation dictionary, EVM bytecode and ABI definitions are stored in the ``
Process Repository''.

The Caterpillar compiler produces three smart contracts from an input model 
if it is flat (i.e., contains no subprocesses): the workflow, worklist, and 
factory contracts. The first of them implements the data and 
control flow perspectives; the data perspective is embedded as part of the control flow 
implementation although.
The second contract, the worklist, handles the execution of workitems by 
stakeholders named and serves to send/receive the process data. 
The third contract, the factory, provides a default mechanism to create 
instances of the process. As stated above, service tasks involve 
interactions with information systems running outside of the blockchain, and 
require other smart contracts similar to worklists. Additionally, some 
modeling elements such as multi-instance activities and call activities are 
implemented in separate contracts. Note that, if a process contains at least 
one subprocess (i.e. embedded or linked to some call activity), then a 
relation parent-child is implicitly established. Considering that each 
subprocess can in turn have a set of children, the subprocesses define a hierarchy.

 \subsubsection{Deploying Smart Contracts}
 
\begin{figure}[hbtp]
 	\centering
 	\includegraphics[width=0.9\textwidth]{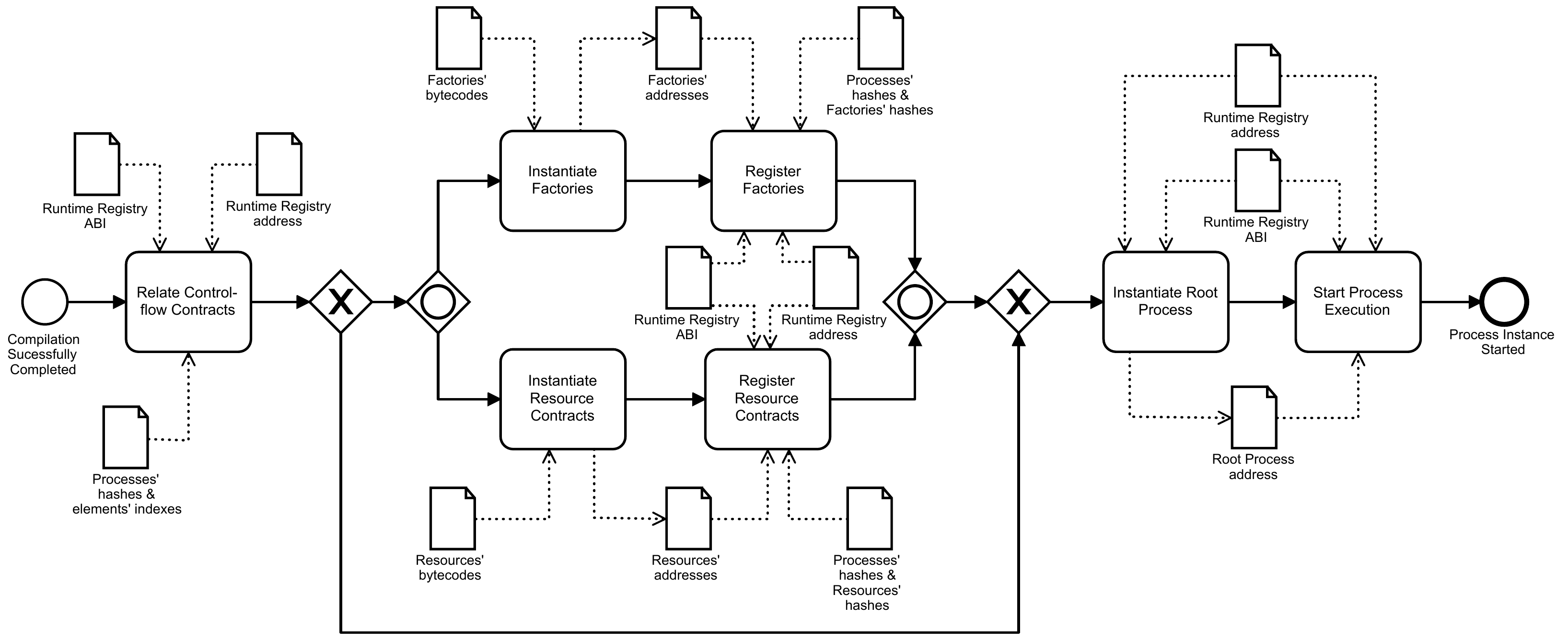}
 	\caption{\label{fig:deployment-process} Caterpillar's instantiation 
process. }
 \end{figure}

 Once the compilation process finishes with the resulting metadata stored in 
the repository, the root contract can be instantiated from the ``Deployment 
Mediator''. Note that, from this point, the contracts' identifiers are the 
hashes produced by the repository which also serves as the key to access the 
compilation metadata. \autoref{fig:deployment-process} shows the steps 
performed by Caterpillar when deploying a contract before creating the first 
instance. Initially, any parent-child relations are updated in the registry. 
Next, for any contract in the process hierarchy the corresponding factories 
and resources (i.e. worklists and services) are instantiated and associated 
accordingly in the registry. Note that instantiating a contract off-chain 
requires the bytecode produced by the solidity compiler. On the other hand, 
calling or invoking a function requires the contract ABI in conjunction with the 
address of a running instance. Finally, the root process is instantiated and 
subsequently started. In contrast to factories and resources that must be instantiated 
explicitly\footnote{There is also possible to register given instances of a 
factories/resources created by another runtime component.}, the workflow 
contract is instantiated through the ``Runtime Registry'' which also updates itself
with the newly created addresses. 
 
 As hinted in \autoref{fig:deployment-process}, most of the steps are optional 
as Caterpillar allows lazy operations. In other words, registrations must be 
performed before elements involving an interaction are reached in the control 
flow, but there is no specific moment to register such information. For 
example, in the process model presented in \autoref{fig:running} it is mandatory 
to register the contract produced for the process {\sc Order to Cash} as root 
to allow the execution of any of its tasks by an off-chain component. Besides, the factory is 
required to create the instance and starting the execution, upon which the 
user task {\tt Submit PO} is reached in the control flow. 
To execute that task, the corresponding worklist is also required. The 
contracts related to the call activity {\sc Goods Shipment} as well as the 
multi-instance {\sc Carrier Selection} can be registered at any moment before 
reaching the respective elements in the control flow. However, it is advisable to 
register any involved interaction before instantiating a contract to avoid 
unexpected runtime errors resulting from reaching unlinked elements during 
the execution.
 
\subsubsection{Querying Process State and Executing Tasks}
 
In order to inform a user which activities she can perform, we need to query the process state -- more specifically, we need to derive the set of started user tasks from the information in the smart contracts.
 Once a process model has been deployed into Caterpillar's runtime for execution, 
any instance of it evolves from its initial state, executing a subset of the 
underlying activities until no activity is found to be a candidate for 
execution or currently being executed. 
Following this intuition, the relevant part of the state of a process instance at a given 
point in time can be identified with the subset of activities that are 
in the state \emph{started}. Note that any activity that is \emph{enabled} during the execution is \emph{automatically started} by Caterpillar, thus simplifying the querying task.
Furthermore, a process instance may give 
rise to more than one contract instance, e.g. from a multi-instance activity or a 
call activity. Nonetheless, it is the address of the instance contract 
related to the root process which would be used for querying the state of the 
process instance as a whole. To that end, Caterpillar traverses the hierarchy 
of smart contracts implementing the behavior of a process instance and 
collects the set of started activities therein.
It is easy to see that 
the hierarchy always corresponds to a tree. Henceforth, querying process 
instance state can be done via traversal of such a tree of contracts.
 
 At any time, an external entity can query the state of a given process 
instance, i.e. the set of activities that are currently active (or executing)
for that particular instance, through the ``Execution Monitor''. Algorithm~
\ref{algo:proc:state} illustrates Caterpillar's off-chain runtime approach to 
querying the state of a process instance. Given the address associated with a 
process instance, the algorithm performs a depth-first traversal over the 
tree representing the hierarchy of contracts, collecting the information of the 
started elements.
 
 \begin{algorithm}[htp]
 	\begin{algorithmic}[1]
 		\Function{instanceStateFor}{process\_address, registry\_address, registry\_ABI}
 		
 		\State runtime\_registry = Contract.at(registry\_address, registry\_ABI)
 		\State workitems, service\_tasks, PENDING $\gets$ $\emptyset$, $\emptyset$, $\emptyset$
 		\State \Call{push}{PENDING, process\_address}	
 		
 		\While {PENDING $\neq \emptyset$}
 		\State instance\_address $\gets$ \Call{pop}{PENDING}
 		
 		\State process\_hash = runtime\_registry.findHashFor(instance\_address, 'PROCESS-CONTRACT')
 		\State dictionary $\gets$ \Call{findDictionary}{Repository, process\_hash}
 		
 		\State instance\_contract $\gets$ \Call{findContractFor}{runtime\_registry, instance\_address, 'PROCESS-CONTRACT'}
 		
 		\State worklist\_address $\gets$ instance\_contract.findWorklist()
 		\State worklist\_contract $\gets$ \Call{findContractFor}{runtime\_registry, worklist\_address, 'WORKLIST-CONTRACT'}
 		
 		\For {element\_index $\in$ instance\_contract.startedActivities()}
 		\Case [dictionary.type\_of(element\_index )]
 		\CaseItem[WORKITEM]
 		\State workitems $\gets$ workitems $\cup$ worklist\_contract.workitemsFor(element\_index, instance\_address)
 		\CaseItem[SERVICE]
 		\State service\_tasks $\gets$ service\_tasks $\cup$ \{(instance\_address, element\_index) \}
 		\CaseItem[SEPARATE\_INSTANCE]
 		\For{subinstance\_address $\in$ instance\_contract.findStartedInstances(element\_index ) }
 		\State \Call{push}{PENDING, subinstance\_address}
 		\EndFor
 		\EndCase
 		\EndFor
 		\EndWhile
 		\State \Return (workitems, service\_tasks)
 		\EndFunction
 		
 		\Statex
 		
 		\Function{findContractFor}{runtime\_registry, instance\_address, contract\_type}
 		\State element\_hash $\gets$ runtime\_registry.findHashFor(instance\_address, contract\_type)
 		\State contract\_abi $\gets$ \Call{findABI}{Repository, element\_hash}
 		\State \Return Contract.at(instance\_address, contract\_abi)
 		\EndFunction

 	\end{algorithmic}
 	\caption{\label{algo:proc:state}Querying process instance state}
 \end{algorithm}
 
 Initially, the input address, i.e. process\_address, is pushed into the 
stack represented by the variable PENDING (line 3) before entering the while 
loop. Besides, the address when running the ``Runtime Registry'', as well as 
its ABI, are required to invoke some functions and then provided as input. In 
line 2, the algorithm calls the function ``Contract.at'' to instantiate a 
contract wrapper (e.g. contract object for solidity contracts according to the 
vocabulary used by the web3.js JavaScript library) which will be later used to 
call functions of the solidity contract running on the blockchain. Note that 
a similar approach is used to recover wrappers from processes and worklists, 
in the function named ``findContractFor'' in lines 27-31. However, in this 
case, firstly it is necessary to recover the hash and ABI from the registry and 
repository respectively.
 
 The traversal is implemented as a while loop (lines 5-24). At each 
iteration, one contract address is processed. In lines 7-8 the algorithm 
retrieves the compilation dictionary which will later be used for determining 
the type of process element being processed. The address of the associated 
worklist is retrieved in line 10, and used later to obtain the worklist wrapper. 
The ``for'' loop in lines 12-24 iterates over the set of started elements that 
Caterpillar on-chain runtime reports for a given address, using the 
corresponding contract wrapper\footnote{As described later in \autoref{sect:compilation}, the set of started 
activities is represented with an integer, manipulated as a bit set and the 
line 12 in the algorithm conceptually captures the iteration over such set. 
Henceforth, each BPMN element in that set is represented by a bit mask, which 
we refer to as element\_index in the algorithm.}.
 
 For the Caterpillar's off-chain runtime, there are two types of activities 
for which the "started" state is externally visible: those associated with 
workitems (i.e. user tasks, receive tasks and message events) and the service 
tasks. Lines 18-21 will be executed when a BPMN element is associated with a 
separate contract instance, e.g. the element found in the started state is a 
call activity. 
In such cases, the for loop in lines 19-21 iterates over the set of contract 
instances associated with the BPMN element at hand, pushing their 
corresponding contract addresses into PENDING for further processing. 
The algorithm returns two lists, one with the indexes of the workitems of the started user tasks 
and a second list with the started service tasks which are represented by a pair enclosing the task 
index and the corresponding instance address. To 
determine which function is associated with an element, the dictionary must 
be queried.
 
 Querying other useful information such as the deployed process models or the 
addresses of running instances are relatively trivial. Such queries 
are made by the ``Execution Monitor'' by calling the corresponding functions 
in the ``Runtime Registry'' or checking the repository.   
 
 The ``Execution Monitor'' also allows executing activities that are started 
when querying the process status by external actors with the required access 
authorization. Nevertheless, such execution must be started from the worklist 
contract which validates the interaction before redirecting the call into the 
workflow contract. Algorithm \ref{algo:exec:task} illustrates the steps needed 
to perform user tasks. To that end, the address of the 
corresponding worklist, the workitem index, as well as the values of the 
input parameters must be provided. Note 
that, although it is not explicit in the Algorithm \ref{algo:proc:state}, the 
information about worklists and parameters required to execute started tasks 
is provided when querying the process status.
 
 \begin{algorithm}[htp]
 	\begin{algorithmic}[1]
 		
 		\Function{executeTask}{worklist\_address, workitem\_identifier, input\_parameters, runtime\_registry}
 		
 		\State worklist\_contract $\gets$ \Call{findContractFor}{runtime\_registry, worklist\_address, 'WORKLIST-CONTRACT'}
 		\State element\_index = worklist\_contract.elementIndexFor(workitem\_identifier)
 		\Statex
 		\State worklist\_hash $\gets$ runtime\_registry.findHashFor(worklist\_address, 'WORKLIST-CONTRACT')
 		\State dictionary $\gets$ \Call{findDictionary}{Repository, worklist\_hash}
 		\State function\_name $\gets$ dictionary[element\_index].function\_name
 		\Statex
 		\State \Call {ExecuteFunction}{worklist\_contract, function\_name, input\_parameters}
 		
 		\EndFunction
 		
 	\end{algorithmic}
 	\caption{\label{algo:exec:task} Executing a task by an external actor}
 \end{algorithm}
 
 Initially, in Algorithm \ref{algo:proc:state} the function ``findContractFor'' 
in lines 27-31 retrieves 
the interface for the worklist contract used later to find the index 
associated with the corresponding BPMN element, i.e. the user task. Next, the 
compilation dictionary provides the name of the function to be executed in 
the worklist, as shown in line 7. In this case, as the function name is just 
a ``String'', the execution occurs without invoking the function on the contract 
wrapper. Instead, a sort of interface (e.g. provided by web3.js JavaScript 
library) is used to resolve such execution.

 Finally, given that interactions with Ethereum are executed 
asynchronously, the ``Event Monitor'' listens for low-level events. This 
component processes the Ethereum event log to determine when a notification 
needs to be pushed to the ``Execution Monitor'', as well as any external 
application, e.g. when a task has completed or when a new workitem is executed.

\subsection{Web Portal}
 
The Caterpillar Web Portal exposes the functionality of the off-chain runtime 
component to end users (e.g.\ process administrators and process workers) via 
a form-based user interface. The Web portal is structured into three panels: 
``Modeling'', ``Configuration'' and ``Execution'' (cf. \autoref{fig:architecture}).
 
The ``Modeling panel'' allows the user to draw the BPMN models that are 
deployed later into the blockchain. Besides, it is possible to import and 
edit models produced by another tool if they comply with the BPMN standard. In 
both cases, the models are typically enriched with the Solidity snippets which 
later are embedded into the smart contracts produced. Note that, in case of 
errors in such solidity code, the compilation fails; after fixing the errors, the modeler can try to compile the model again. As indicated in \autoref{fig:architecture}, the models created in the ``Modeling panel'' are deployed 
to the blockchain through the ``Deployment mediator'' if the compilation 
succeeds.
 
The ``Configuration panel'' supports introducing new relations or updating 
the links defined for a process model already deployed on the blockchain. For 
example, it is possible to import and instantiate a smart contract of a 
worklist produced not by Caterpillar (but following the structure expected by 
the tool) that are associated later with a deployed process. This 
component is also the entry point to deploy and instantiate the contracts 
which implement the interactions with the service tasks. Besides, 
changing the links of a process to another, e.g. through a call activity, is 
also supported. In other words, the ``Configuration panel'' provides a set of 
functionalities to introduce and update the information in the ``Runtime 
Registry'' at any time during the process execution.
 
The ``Execution Panel'' interacts with the ``Execution monitor'' to retrieve 
all the information about deployed models, running instances and also allow 
executing tasks by stakeholders. For example, a user can access the list of 
deployed processes to choose the desired one. Then, all the information 
stored in the ``Distributed repository'' can be checked as well as the list 
of running instances can be retrieved, in this case from the ``Runtime 
Registry''. Besides, given one instance of a selected process, a user can 
visualize its state. Moreover, the started user tasks can be executed, more 
specifically, the ``Execution Panel'' will generate a Web form to let users 
provide input data to a process instance which will be validated by the 
corresponding worklist. Finally, the ``Execution panel'' receives 
notifications from the ``Event Monitor'' to keep track of the transactions 
included in the blockchain and update the visualization accordingly.
 

\section{Compiling BPMN into Solidity Smart Contracts} \label{sect:compilation}

\begin{figure}[hbtp]
 	\centering
 	\includegraphics[width=0.62\textwidth]{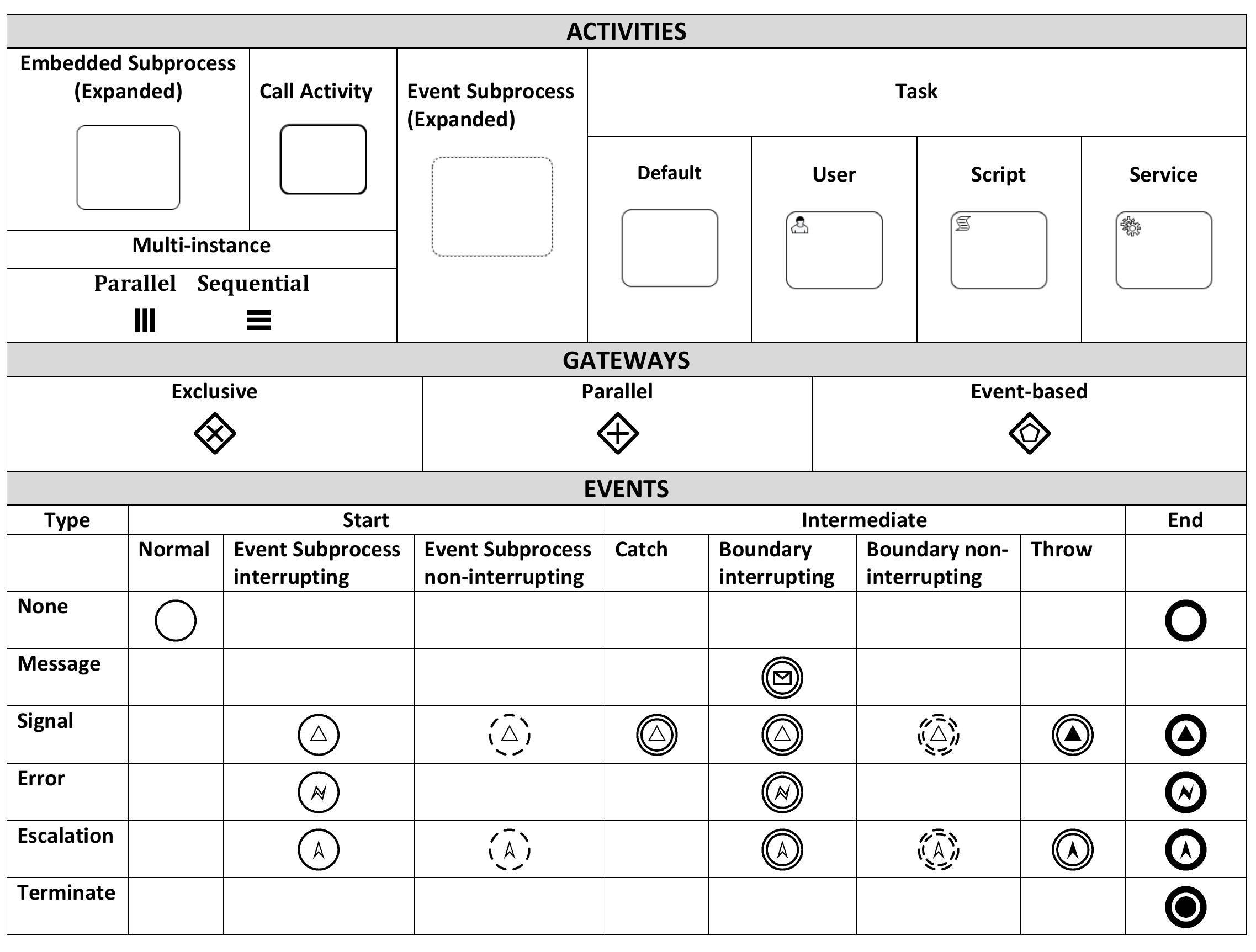}
 	\caption{\label{fig:bpmn-elements} BPMN elements supported by Caterpillar.}
\end{figure}

The Caterpillar BPMN compiler takes as input a BPMN process model annotated with Solidity code snippets. The input BPMN model may contain user tasks, service tasks, script tasks (with Solidity scripts), gateways (exclusive, parallel and event-based), events (default, terminate, message, signal, error and escalation), call-activities, embedded subprocesses, event-subprocesses and multi-instance activities (parallel and sequential), as summarized in \autoref{fig:bpmn-elements}. 
It also supports message events attached to the boundary of an activity (i.e.\ interrupting and non-interrupting message boundary events).

\vspace*{-5mm}
\vspace*{-5mm} 
 
Below, we discuss how Caterpillar generates smart contacts from a BPMN process model. We start by discussing the handling of variables and interactions with external resources via user and service tasks. We then discuss how the control-flow (sequence flows and gateways), subprocesses, and events are encoded in the generated smart contracts. 

\subsection{Process variables and external resources} 
\label{sect:external}

\autoref{lst:excerpt1} shows an excerpt of the smart contract that Caterpillar generates from the order-to-cash process in \autoref{fig:running}. This contract is called the \emph{main contract} or the \emph{process contract}. Lines 6-10 define the variables of the process. These variables are defined in the global documentation of the model, from where Caterpillar copies them into the smart contract. 
 
\begin{lstlisting}[float, language=Solidity, caption=Example of data flow and resource management in the solidity contract of the process Order to cash.,label={lst:excerpt1},frame=tb]
contract OrderToCashProcess {
	// == RESOURCE MANAGEMENT ==
	address worklist;
	
	// == DATA FLOW PERSPECTIVE ==
	enum POStatus {PENDING, ACCEPTED, REJECTED, CANCELED, CLOSED}
	bytes32 sku;
	uint quantity;
	uint price;
	POStatus status;
	...
	
	function ValidatePO_Complete(POStatus decision) external {
		require(msg.sender == address(worklist) && /* Control flow validations */);
		require(decision == POStatus.ACCEPTED || decision == POStatus.REJECTED);
		status = decision;
		// Continues with the execution of next elements in the process flow
	}
	
	...	
}
\end{lstlisting}
 
 Lines 13-18 outline the function {\tt ValidatePO\_Complete}, which is generated by Caterpillar from the ``Validate PO'' user task in the running example. Tasks designed to interact with external resources (user and service tasks) may read and write from/to the variables in the process contract. This data mapping is specified in the task's specification using the syntax {\tt <Data\_to\_export> : \tt <Data\_to\_import> --> <Operations\_to\_perform>}. The {\tt <Data\_to\_export>} section defines which variables are read by the task from the process contract (i.e.\ input parameters of the task). The {\tt <Data\_to\_import>} specifies the output parameters of the task, i.e.\ the data that the task obtains from the external resource.  The {\tt Operations\_to\_perform} section contains a set of Solidity operations to map the output parameters to the variables of the process. For example, the {\tt Validate PO} task takes as input the stock keeping unit (sku), the quantity and the price per unit. It returns the decision that the user makes on the PO (to accept it or reject it). This output parameter is type-checked and then copied to the ``status'' variable. The corresponding task specification is:
 \begin{center}
 	{\tt (bytes32 sku, uint quantity, uint price) : (POStatus decision) ->}
 	
 	{\tt \{ require(decision == POStatus.ACCEPTED || decision == POStatus.REJECTED); status = decision; \}} \footnote{The Solidity statement {\tt require} is used as a precondition at the beginning of a function to check if the underlying transaction would fail or not.}
 \end{center}




A smart contract has no way of calling an external resource directly. To cope with this limitation, we publish Solidity events on a log that is stored inside each full blockchain node, which external programs can read. 
The external programs and their users react to the events and submit responses in the form of transactions to the blockchain.
A  software that follows this style of interaction is often referred to as an oracle~\cite{ethereumoracles}. 
An oracle basically serves as a proxy to mediate between smart contracts and external applications, and often includes another contract dedicated to it. 
\autoref{lst:worklist} sketches the contract {\tt OrderToCashWorklist}, which acts as a proxy to handle the interactions generated by user tasks in the order-to-cash process. 
Although any BPMN element that can be triggered by a user generates the corresponding functions in the contract, below we only discuss the methods of task {\tt Validate PO} to illustrate the approach.
 
For every user task, Caterpillar generates two methods in the worklist contract, i.e, {\tt ValidatePO\_Start} and {\tt ValidatePO} in {\tt OrderToCashWorklist}. In the process contract, the task is implemented as one function, {\tt ValidatePO\_Complete} in \autoref{lst:excerpt1}. To generate these functions, Caterpillar uses the following approach:
 
 \begin{itemize}
 	\item {\tt <element\ name>\_Start(<Data\_to\_export>)}. This function must be triggered internally when the corresponding element is reached during the execution flow in the process contract (see control flow \autoref{sect:controlflow} for further details). The method includes as parameters the set of variables annotated as {\tt Data\_to\_export} in the corresponding BPMN element in the process model. The worklist contract includes the corresponding workitem that is stored in a dynamic array, called {\tt workitems}. A workitem contains the address of the process contract that made the requests and the index that identifies the element in the process contract. Workitems would be accessible by the external resource and would be retrieved from an external application when required. The input parameters and the index of the workitem in the dynamic array that works as the workitem identifier are stored in a Solidity event on the Ethereum log, which is also visible to external applications.
 	\item {\tt <element\ name>(<Data\_to\_import>)}. This function should be called by the external resource to provide the information required to continue with the execution of the element. Then, the worklist sends the data to the process contract and marks the workitem as \emph{completed}, by setting the address to 0. 
The external resource must fulfill any access control policies defined for the process. In the example displayed in \autoref{lst:worklist}, any user is allowed to access any function.
 	\item {\tt <element\ name>\_Complete(<Data\_to\_import>)}. This function, implemented by the process contract, is invoked by the associated worklist. The function first checks that the blockchain address performing the call matches the worklist's address, which is stored as a global variable in the process contract (line 3 in \autoref{lst:excerpt1}). Next, the function calls {\tt <Operations\_to\_perform>} to update the process variables. The task is then marked as ``completed'' and the state of the process instance is updated. If a second call arrives from the worklist, it is rejected. 	
\end{itemize}
 
 
 	\begin{lstlisting}[float, language=Solidity, caption=Example Worklist and interaction interface of the process Order to cash  contract.,label={lst:worklist},frame=tb]
 	
contract OrderToCashWorklist {
 	struct Workitem {
 		address instanceAddress;
 		uint elementIndex;
 	}
 	
 	Workitem[] public workitems;
 	
 	event ValidatePO_Requested(uint workitemId, bytes32 sku, uint quantity, uint price);
 	
 	function ValidatePO_Start(bytes32 sku, uint quantity, uint price) external {
 		workitems.push(Workitem(msg.sender, 2));
 		ValidatePO_Requested(workitems.length - 1, sku, quantity, price);
 	}
 	
 	function ValidatePO(uint workitemId, uint decision) external {
 		require(workitemId < workitems.length);
 		require(workitems[workitemId].elementIndex == uint(2));
 		require(workitems[workitemId].instanceAddress != address(0)); 
 		WorklistInterface(workitems[workitemId].instanceAddress).ValidatePO_Complete(decision);
 		workitems[workitemId].instanceAddress = address(0);
 	}
 	...
}
 	
contract WorklistInterface {
 	// == FUNCTIONS IN WORKLIST CONTRACT ==
 	function ValidatePO_Start(bytes32 sku, uint quantity, uint price) external;
 	function ValidatePO(uint workitemId, uint decision) external;
 	// == FUNCTION IN MAIN CONTRACT ==
 	function ValidatePO_Complete(POStatus decision) external;	
 	...
}
 	\end{lstlisting}
 
 
To achieve modularity, Caterpillar generates an interface to handle the function calls between the worklist contract and the process-related contracts. Thanks to this interface, we avoid direct interaction between Solidity contracts, which would lead to higher gas consumption during deployment\footnote{https://medium.com/daox/avoiding-out-of-gas-error-in-large-ethereum-smart-contracts-18961b1fc0c6}. Instead, we create a third contract ({\tt WorklistInterface} in lines 27-34 in \autoref{lst:worklist}), which mediates all calls to the worklist contract. Accordingly, all the function calls are of the form {\tt Contract\_Interface(<contract\_address>).<function\_name>(<function\_parameters>)} (cf.\ line 21 of \autoref{lst:worklist})).
 
Service tasks are handled in a similar way as user tasks. 
The main difference is that for service tasks, the API  of the oracle contract is specified in the process model, whereas for user tasks, the oracle is implemented by the predefined worklist contract. 


\subsection{Control Flow Perspective}\label{sect:controlflow}

To encode the execution status of each element in a BPMN model, we rely on the following classification of BPMN elements: 
 \begin{description}
 	\item[External:] Elements involving external resources, i.e., user tasks, receive tasks, message catching events and service tasks.
 	\item[Reusable:] Elements whose execution instantiates a process represented by an external process model (not necessarily provided) or which trigger a sub-routine specified within the current process model. This class includes call activities, multi-instance activities, non-interrupting boundary events, and event subprocesses.
 	\item[Internal:] All other elements: script tasks, gateways (exclusive, parallel and event-based), intermediate and end events (default, terminate, message, signal, error, and escalation). We also put embedded (single-instance) subprocesses in this category since they can be in-lined into the parent process and hence do not trigger a separate sub-routine.
 \end{description}
  
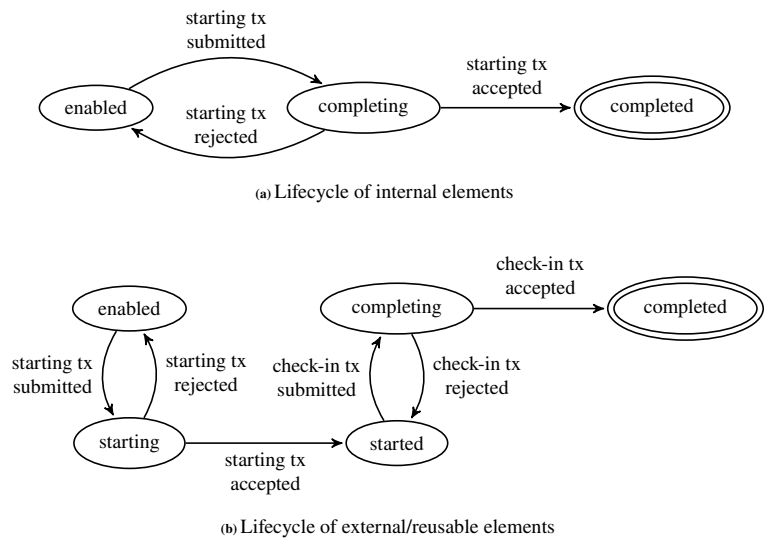
\begin{figure}[hbtp]
  \centering
		
  \begin{subfigure}[b]{0.9\textwidth}
	\centering
	 \begin{tikzpicture}[->,>=stealth',shorten >=1pt,auto,node distance=1cm,
		semithick, font=\footnotesize]
		\matrix[column sep=1.75cm,row sep=1cm]{
			\node[ellipse,draw] (a) {enabled}; & \node[ellipse,draw] (d) {completing}; & \node[ellipse,draw,inner sep=5pt] (e) {completed};\\
		};
		\node[ellipse,draw] at (e) {\phantom{completed}};
		
		\path 	
		(a) edge[bend left] node [right,above,align=center] {starting tx\\ submitted} (d)
		(d) edge[bend left] node[right,above,align=center] {starting tx\\ rejected} (a)
		(d) edge node [above,align=center] {starting tx\\ accepted} (e)
		;
	\end{tikzpicture}
	\caption{\label{subfig:internal}\footnotesize Lifecycle of internal elements}
  \end{subfigure}	
	\medskip
		
  \begin{subfigure}[b]{0.9\textwidth}
 	\centering
    \begin{tikzpicture}[->,>=stealth',shorten >=1pt,auto,node distance=1cm,
	    semithick, font=\footnotesize]
	    \matrix[column sep=1.75cm,row sep=1cm]{
	    	\node[ellipse,draw] (a) {enabled}; & \node[ellipse,draw] (d) {completing}; & \node[ellipse,draw,inner sep=5pt] (e) {completed};\\
	    	\node[ellipse,draw] (b) {starting}; & \node[ellipse,draw] (c) {started};\\
	    };
	    \node[ellipse,draw] at (e) {\phantom{completed}};
	    \node[above=6mm of a] (start) {};
	    
	    \path 
	    (a) edge[bend right] node [left,align=center] {starting tx\\ submitted} (b)
	    (b) edge[bend right] node [left,right,align=center] {starting tx\\ rejected} (a)
	    (b) edge node[below,align=center] {starting tx\\ accepted} (c)
	    
	    (c) edge[bend left] node [right,left,align=center] {check-in tx\\ submitted} (d)
	    (d) edge[bend left] node[right,align=center] {check-in tx\\ rejected} (c)
	    (d) edge node [above,align=center] {check-in tx\\ accepted} (e)
	    ;
    \end{tikzpicture}
 	\caption{\label{subfig:external}\footnotesize Lifecycle of external/reusable elements}
 \end{subfigure}
 	\caption{Lifecycle of BPMN elements in Caterpillar: (\subref{subfig:external}) external/reusable (\subref{subfig:internal}) internal}
   \label{fig:lifecycle} 
\end{figure}

The lifecycle of a BPMN element in Caterpillar depends on its class as depicted in \autoref{fig:lifecycle}. Commonly, a BPMN element becomes enabled when a token is present on its incoming edges (one of them in the case of exclusive join gateways, all of them otherwise). Boundary events are enabled if they are attached to a started activity. Finally, an event subprocess becomes enabled if it is included in a process/subprocess which contains, at least, one element enabled or started. The execution of an enabled element consumes the incoming tokens, generating new ones on its outgoing edges as described further below.
  
The lifecycle of internal elements is the simplest. As there are no external interactions, the element is automatically executed the moment it becomes enabled. 
This happens when a previous element has been executed, and within the scope of the transaction that executed the previous element. 
For an example, consider the process model in \autoref{fig:running} and the contracts in \autoref{lst:excerpt1}, \autoref{lst:worklist}. 
When a transaction calls the function {\tt ValidatePO}, the transaction executes the respective workitem function, which calls the workflow function {\tt ValidatePO\_complete}; this function in turn executes the token flow and calls the function corresponding to the data-based XOR split, cf.\ \autoref{fig:running}; depending on the decision, further functions will be called, all within the scope of the single transaction to {\tt ValidatePO}. In the lifecycle, we refer to this transaction as the ``starting tx''. 
The state ``completing'' is used to capture the validation and inclusion (or not) of  transactions on the blockchain (since transactions can be rejected).
The state becomes ``completed'' when this transaction is included in the blockchain. 
Caterpillar performs every internal operation in a single transaction until reaching a step where only calls to functions in other contracts are pending.

The execution of an external element requires an interaction with an oracle as discussed in~\autoref{sect:external}. When an external element is enabled during the execution from a ``starting tx'' as above, the oracle contract (e.g.\ worklist) is invoked and the element is in state starting/started. 
Unlike internal elements, an external element requires a separate transaction, called ``check-in tx'' in the lifecycle, to proceed.
Eventually the oracle makes the corresponding response call, and the element is then marked as ``completed''. 

In the running example, once the task {\tt ValidatePO} is enabled, the contract {\tt OrderToCashProcess} invokes function {\tt ValidatePO\_Start} of {\tt OrderToCashWorklist}, which corresponds to the ``start'' transaction in the lifecycle. Consequently, the task is checked-in when an authorized user triggers the function {\tt ValidatePO} in {\tt OrderToCashWorklist}. Then, the worklist invokes {\tt ValidatePO\_Complete} in {\tt OrderToCashContract}, which in turn generates a new token in the outgoing sequence flow of the task, and the execution continues as outlined above.
 
Reusable elements also involve interactions with external contracts encapsulating the behavior of another process. Therefore, the lifecycle is similar to external entities. Once a reusable element is enabled, it creates as many new instances as specified in the corresponding contract. In this case, the element is considered started while any of the instances of the external contracts are running on the blockchain. 
Intuitively, the element completes when no entity is enabled or started in any of the instances. The simplest case of a reusable element is the call activity. Here, the metadata required to instantiate the contracts of the process associated with the call activity must be provided. Otherwise, an execution error occurs when the call activity becomes enabled. Caterpillar creates the contracts of the other reusable elements as described in \autoref{sect:reusable}.
 
The occurrence of an interaction with an external/reusable entity triggers changes in the state of the process instance. Such updates correspond to recomputing the new distribution of tokens over the set of elements in the model and, as consequence of this, executing internal elements or starting some others that also require interaction with other external entities.

\medskip
The control flow perspective of a process is implemented by simulating a token game as specified in the BPMN standard. When a process instance is created, a token is generated by the start event, which traverses sequence flows in the model until reaching the end event(s). To simulate the token game, we assign sequential indexes (starting from 1) to each node in the BPMN model. We use the index 0 as the identifier of the process. For the flow arcs, we follow the same approach but starting with zero. 
Following the approach of our earlier work\cite{GarciaPDW17}, we assume that the model is 1-safe, meaning that it is designed in a way that at most one token is present on a sequence flow at any time. This property allows us to use a bit array, to encode the distribution of tokens in a given state of a process instance. These bit arrays are encoded as 256-bits unsigned integers, which is the default word size in the Ethereum Virtual Machine.
 
Each process contract generated by Caterpillar contains two integer variables called {\tt marking} and {\tt startedActivities} to encode the current state of a process instance. Variable {\tt marking} is a bit-array encoding the distribution of tokens across the sequence flows of the process model. Each sequence flow is associated with one bit in this variable: 1 if the sequence flow has a token, 0 otherwise. Variable {\tt startedActivities} encodes the set of triggered external/reusable elements. Each bit in {\tt startedActivities} corresponds to an external/reusable element. Moreover, a dynamic array named {\tt subInstanceAddresses} stores the addresses of every instance created by each reusable element. Variable {\tt subInstanceStartedIndexes} is a bit array that tracks which of these reusable elements are started. 
Finally, attribute {\tt worklist} keeps the address of the contract that handles user tasks.\footnote{In the current implementation, all bit vectors limit the respective content to 256 (e.g., sequence flows). While this limit could easily be lifted, doing so would likely increase the gas cost.}
 
 We use bit-wise operations to handle all the queries/updates on the process state. To check if an element is enabled or started, the bit-wise \textit{AND} allows testing set inclusion. The bit-wise \textit{OR} provides a method to encode the set union as an integer. This operator is used to append tokens in the {\tt marking}, to group the indexes of the incoming edges of an exclusive gateway or the elements contained in a subprocess, among others. Finally, the combination of \textit{NOT} and \textit{AND} serves to remove tokens/elements from the variables {\tt marking}/{\tt startedActivities}.
 
The control flow implementation is illustrated in \autoref{lst:step}, which complements \autoref{lst:excerpt1} with other attributes and operations required to manage the control-flow perspective.
Consider the function {\tt ValidatePO\_Complete} shown in lines 14-18. 
This function is called by the worklist when a user performs the task {\tt ValidatePO}. 
Line 15 requires {\tt startedActivities \& uint(4) != 0}, meaning that the activity corresponding to 4, i.e., {\tt ValidatePO\_Complete}, is started.
4 here is the decimal representation of the binary number 100, i.e., the bit in the third-last position of the bit array, or index position 2.
Note that, as part of a previous transaction, the {\tt if} statement in lines 24-29 must start the interaction with the worklist when the task became enabled (i.e., condition {\tt tmpMarking \& uint(2) != 0} is {\tt true}). As result, the token on the incoming arc is removed, {\tt tmpMarking \&= uint(~2)}, and the task is started, {\tt tmpStartedActivities |= uint(4)}. 
The numbers are basically the indexes assigned to the corresponding arcs and tasks, respectively, when compiling the model into Solidity. Note that we need to cast every bitmask to avoid overflows in the operations because an integer literal is mapped in solidity to \textit{uint8} instead of \textit{uint256}.

\begin{lstlisting}[float, language=Solidity, caption=Example of a solidity contract to execute the control flow perspective.,label={lst:step},frame=tb]
	contract OrderToCashProcess {
		...
		RuntimeRegistry private registry;
		
		uint private marking = 1;
		uint private startedActivities = 0;
		
		address private parent = 0;
		uint private instanceIndex;
		
		address[] private subInstanceAddresses;
		mapping(uint => uint) private subInstanceStartedIndexes;
		
		function ValidatePO_Complete(POStatus decision) external {
			require(/* Resource validations */ && startedActivities & uint(4) != 0);
			// <Operations_to_perform> --> Data perspective updates
			step(marking | uint(4), startedActivities & uint(~4));
		}
		
		function step(uint tmpMarking, uint tmpStartedActivities) internal {
			while (true) {
			    ...
			    // User Task (external resource interaction)
				if (tmpMarking & uint(2) != 0) {
					WorklistInterface(worklist).ValidatePO_Start(sku, quantity, price);
					tmpMarking &= uint(~2);
					tmpStartedActivities |= uint(4);
					continue;
				}
				// XOR Gateway (internal element)
				if (tmpMarking & uint(4) != 0) {
					tmpMarking &= uint(~4);
					if (poStatus == POStatus.ACCEPTED)               
						tmpMarking |= uint(16);
					else                 
						tmpMarking |= uint(8);
					continue;
				}
				// Call Activity (reusable element)
				if (tmpMarking & uint(16) != 0) {
					tmpMarking &= uint(~16);
					address child = registry.newInstanceFor(uint(3), this);
					uint index = subInstanceAddresses.length;
					subInstanceAddresses.push(child);
					subInstanceStartedIndexes[uint(3)] |= (uint(1) << index);
					AbstractProcess(child).setInstanceIndex(index);
					tmpStartedActivities |= uint(8);
					continue;
				}
				...
				break;
			}
			marking = tmpMarking;
			startedActivities = tmpStartedActivities;
		}
		...
	}
\end{lstlisting}

The computation of the new process state happens inside a function called {\tt step} displayed in lines 20-55 of \autoref{lst:step}. 
The function {\tt step} is internal, which means that external actors cannot call it. However, each time an external entity's function call updates the state of the process contract, the function {\tt step} is invoked.

To illustrate how the {\tt step} function works, consider again function {\tt ValidatePO\_Complete} shown in lines 14-18. In line 17 the function {\tt step} is called to update the process state. When making this call, the outgoing edge is activated by changing the {\tt marking}, and {\tt ValidatePO} is marked as completed by changing {\tt startedActivites}.
The function {\tt step} receives as input a copy of these changed values of {\tt marking} and {\tt startedActivities}. It is a common practice in Solidity to copy the values of contract variables into local ones, as a way to reduce the number of write operations over contract variables which are costly\footnote{The EVM has three areas to store items. (1) The \textit{storage}, where all the contract state variables are located, and it is costly to use. (2) The \textit{memory} to hold temporary values which is cheaper. (3) The \textit{stack} to hold small local variables that is almost free, but it allows a limited amount of values.}. Later, function {\tt step} identifies the set of BPMN elements that are enabled based on the current marking. Indeed, the function {\tt step} repeatedly executes a sequence of {\tt if} statements to determine whether the current marking enables a given BPMN element. If that is the case, the {\tt step} function starts the execution of the corresponding element. Due to concurrency, it is possible that more than one element gets enabled in a given state. For this reason, the function {\tt step} has to restart the while loop until no more enabled elements are found. It is only at this moment that the newly computed instance state is stored in the contract variables {\tt marking} and {\tt startedActivities}.

Let us to come back to the execution of {\tt ValidatePO} in \autoref{lst:step}. When calling {\tt step} from {\tt ValidatePO\_Complete}, the task is completed after updating the corresponding bit from {\tt tmpStartedActivities}, and adding a new token on the arc with index 2. As a result, the next exclusive gateway is enabled and executed internally as shown in lines 31-38.
The conditions of an exclusive gateway must be boolean expressions encoded in Solidity; they are attached to the outgoing arcs of the gateway in the model, and embedded in the generated code as shown in line 33.       

Finally, we describe how a reusable element is executed. Let us consider that the call activity {\tt GoodsShipment} is enabled, which is handled by the {\tt if} statement in lines 40-49. Checking if the element is enabled and updating the process state is handled analogous to external tasks. However, in this case a new instance of a related smart contract is created (lines 42-46). 
The new instance is created through the registry, which requires the address of the parent and the index assigned to the child when compiling the model. Accordingly, the registry picks the factory to create a new instance of the child from the index provided, which requires that a \textit{parent-child-factory} relation exists in the registry and a factory instance is running on the blockchain. Otherwise, an execution error is thrown. Then, the parent updates variables {\tt subInstanceAddresses} and {\tt subInstanceStartedIndexes} with the new child address and status (lines 41-42). Also, the child updates the attributes {\tt parent} and {\tt instanceIndex}. The former one, as the name suggests, is the address of the parent and is set during deployment through the factory. The latter one keeps the position where the children address was added in the dynamic array of the parent.
Finally, the call activity is marked as ``started''.

\subsection{Subprocesses and Reusable Elements} \label{sect:reusable}

Caterpillar implements three of five types of subprocesses in BPMN:  (1) call activity, which invokes a process defined in a separate process model; (2) embedded subprocess, which invokes a process model embedded inside its parent process; (3) event subprocess, which are like embedded subprocesses but that are triggered by an event.
 
Embedded subprocesses can be inlined in their parent process (and hence are \emph{internal} elements), except when they have a multi-instance marker. In the latter case, Caterpillar generates a separate Solidity contract to encode the multi-instance subprocess. This contract will be instantiated once for each instance of the subprocess.
In the absence of a multi-instance marker, a nested embedded subprocess is inlined inside its closest parent (reusable) subprocess or inside the root process.
  
\begin{figure}[hbtp]
	\centering
	\includegraphics[width=0.5\textwidth]{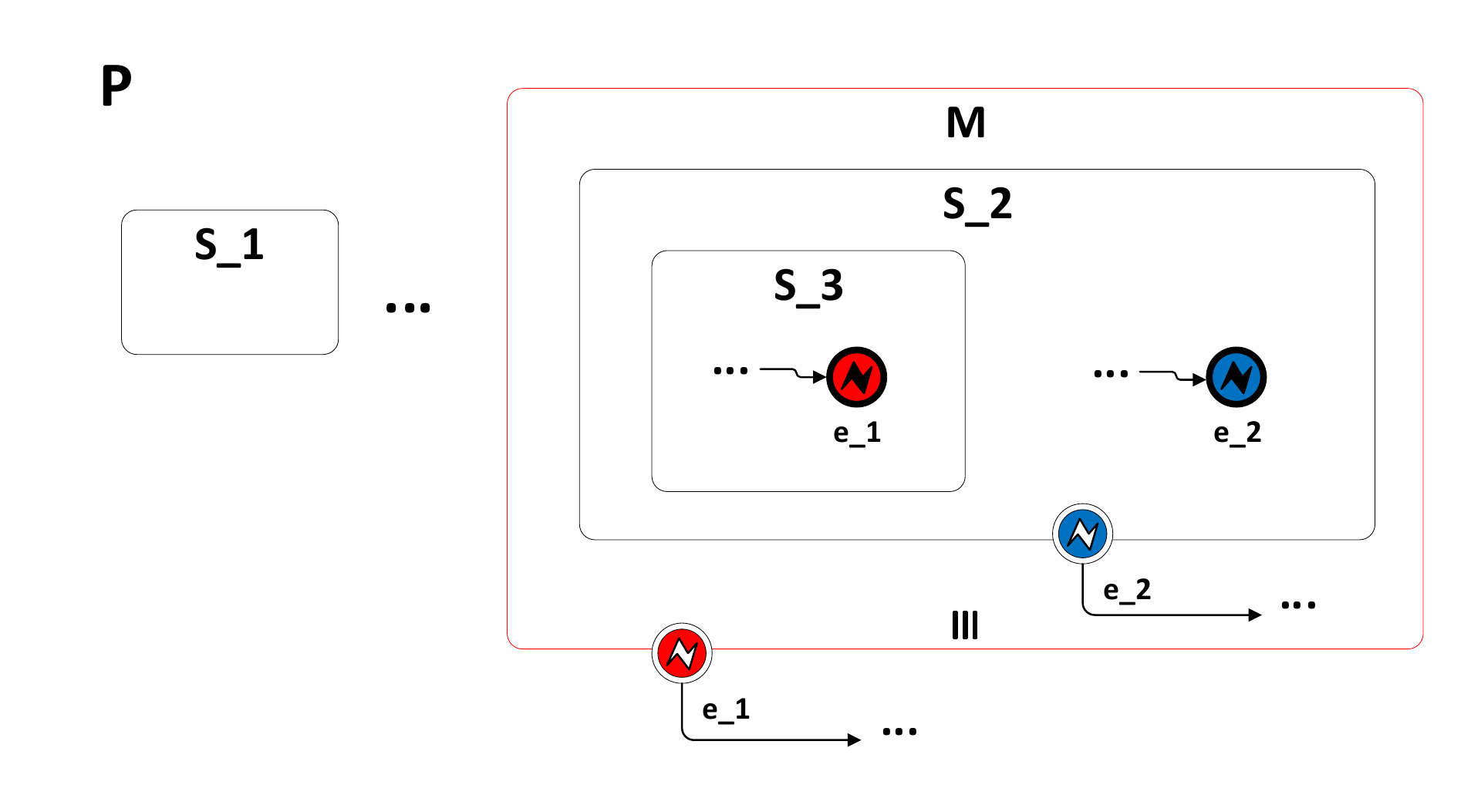}
	\caption{\label{fig:nested-subprocesses} Nested subprocesses with propagation of error events.}
\end{figure}

\autoref{lst:stepProc} provides pseudo-code illustrating possible blocks of instructions in the function {\tt step} to handle embedded subprocesses and multi-instance subprocesses (see \autoref{fig:nested-subprocesses}). Lines 6-9 refers how to start the execution of an internal subprocess, i.e, $S\_1$ in $P$. When the subprocess is enabled in the execution flow, the function {\tt step} automatically executes the start event of the subprocess adding tokens on its outgoing edges. As the elements in $S\_1$ are part of $P$, the execution continues as described in \autoref{sect:controlflow}. Note that the functions ending with \textit{mask} refer to the integer encoding arcs/elements into a bit-array.

 
 	\begin{lstlisting}[float,language=Solidity, caption=Subprocess execution through the step function.,label={lst:stepProc},frame=tb]
 	contract P {
  		...
		function step(uint tmpMarking, uint tmpStartedActivities) internal {
 			while (true) {
      			...
 				if (tmpMarking & incomingEdgesMask(S_1) != 0) {
 					tmpMarking & uint(~incomingEdgeMask(S_1)) | outgoingEdgeMask(startEvent(S_1));
 					continue;
 				}
 				// Parallel Multi-instance of M
 				if(tmpMarking & incomingEdgesMask(M) != 0) {
 					tmpMarking & uint(~incomingEdgesMask(S_1))
 					for (uint i = 0; i < number_of_instances; i++) {
 						// Create a single instance of the contract M. 
 						// Same as call activities, see \autoref{lst:step}, lines 42-46 
 					}
 					tmpStartedActivities |= nodeIndexMask(M)
 					continue;
 				}
 				// Sequential multi-instance of M
				if(tmpMarking & incomingEdgesMask(M) != 0) {
					tmpMarking & uint(~incomingEdgesMask(S_1))
					// Create a single instance of the contract M. 
					// Same as call activities, see \autoref{lst:step}, lines 42-46
					for (uint i = 0; i < number_of_instances - 1; i++)
						subInstanceAddresses.push(0);
					tmpStartedActivities |= nodeIndexMask(M)
					continue;
				}
 				...
 				break;
 			}
 			marking = tmpMarking;
 			startedActivities = tmpStartedActivities;
  		}
 		...
 	}
 	\end{lstlisting}

Although the elements in $M$ are not encoded as functions in the contract of $P$, the instances of $M$ must be started by $P$ when $M$ is enabled in its flow execution. \autoref{lst:stepProc} shows the two possible cases handled by Caterpillar. As illustrated in lines 11-19, if $M$ is a parallel multi-instance, the approach differs to instantiate a call activity only in the {\tt for} loop to create several instances. The subprocess remains started in $P$ while at least one instance of $M$ is running (i.e. at least one instance with {\tt marking} or {\tt startedActivities} different from zero). A default event is propagated to $P$ when any instance of $M$ is finished, thus the marking in $P$ must be updated accordingly. As shown in lines 21-29, sequential multi-instances involve the creation of one instance. Then, the spots required for the next ones are reserved in the dynamic array {\tt subInstanceAddresses}. The next instance is created when the previous one propagates an event notifying its ending. The subprocess remains started until the last instance notifies that it finished. For further details, we fully describe the event propagation among contracts in \autoref{sect:evthand}.

An event subprocess is treated as \emph{internal} when triggered by an interrupting event. Otherwise, it is treated as \emph{reusable} and hence a separate contract is generated for it. This is because non-interrupting events may be triggered several times during the execution of a process. Thus, they are similar to multi-instance subprocesses. Non-interrupting boundary events are special cases of reusable elements. Here, Caterpillar creates a separate contract including the subgraph of elements reachable from the event. As a result, we obtain a new node in the process hierarchy with the same behavior of a non-interrupting event subprocess. 

\subsection{Event Handling} \label{sect:evthand}

Event handling in BPMN refers to the actions of throwing and catching events and the state changes arising thereof.
Some events change the state of the process in a trivial way: They merely move a token from their input to their output sequence flow(s). However, other events lead to their enclosing process instance being terminated and may need to be propagated upwards through the process hierarchy.
The latter include end events in event-subprocesses, terminate and error events, and boundary events attached to an activity. Below we discuss how Caterpillar handles the \emph{event propagation} engendered by such events.

Caterpillar distinguishes four cases of event propagation, when an event $e$ is thrown in a subprocess $S$:
\begin{description}
	\item[Upward:] $S$ propagates $e$ to its parent $P$ in the hierarchy. If $P$ contains a catching event that matches $e$, then it is handled, and the propagation stops. Otherwise, $P$ propagates $e$ to its parent and so on. If $P$ has no parent (root) and no subprocess could catch $e$ on the path from $S$ to $P$, the propagation finishes and, depending on the event type, the running instances are stopped or not.
	\item [Single upward:] $S$ propagates the event to its parent that shall handle it.
	\item [Broadcast:] $S$ propagates $e$ to the root process. Then, the event flows to any started child subprocess that is reachable by traversing the hierarchy from the root. Accordingly, any enabled catching event with the same type of $e$ must handle it when reaching the corresponding subprocess.
	\item [Outside:] $S$ pushes the result of the event to a resource interacting with the process from outside of the hierarchy. 
\end{description}

The propagation always starts when a throwing event is enabled. These type of events are internal, thus executed in the function {\tt step}. 
If the propagation does not traverse any reusable subproceses, the token distribution resulting from the propagation is computed at compilation-time and encoded in the {\tt step} function, which would continue. 
If, on the other hand, the propagation traverses reusable elements, the resulting token distribution cannot be computed at compilation-time. The Pseudo-code in \autoref{lst:stepEvt} illustrates how the propagation of throwing events is handled in the {\tt step} function. The pseudo-code includes, as comments, the operations/analysis to perform when compiling the model, as well as the generated Solidity code. For conciseness, parameters {\tt tmpMarking} and {\tt tmpStartedArtivities} are shortened as {\tt tM} and {\tt tS}, respectively. The instructions to check if an event $e$ is enabled and for removing tokens from its incoming edges is common to all cases (cf.~lines 4-5). The rest of the generated Solidity code is case-dependent.


 	\begin{lstlisting}[float, language=Solidity, caption=Event propagation.,label={lst:stepEvt},frame=tb]
 	  function step(uint tM, uint tA) internal {
 	    while(true) {
        ...
        if (tM & /*incomingEdgesMask(e)*/ != 0) {
          tM &= uint(~/*incomingEdgesMask(e)*/);
/* (1) Case: Upward --> Error or Escalation ------------------------------------------ */
/*       e' = closestLocalCatchingEvent(e) */
/* (1.1) IF NOT exist e' */
		      (tM, tA) = propagateEvent(/*type(e)*/, /*code(e)*/, /*index(e)*/, tM, tA);
/* (1.2) ELSE IF isInterrupting(e') AND isStartEvent(e') 
         C = subprocessEnclosing(e') */
		      (tM, tA) = killProcess(/*index(parent(C))*/, tM, tS);
		      tM |= /* initialMarkingMask(C) */
/* (1.3) ELSE IF isInterrupting(e') AND isBoundaryEvent(e') 
         C = subprocessWhereAttached(e') */
		      (tM, tA) = killProcess(/*index(C)*/, tM, tS);
		      tM |= /*outgoingEdgesMask(e')*/;
/* (1.4) ELSE IF isNotInterrupting(e') */
		      createNewSubprocessInstance(/*index(e')*/);
		      tS |= /*index(e')*/;
/* (2) Case: Single upward --> Default End Event   ----------------------------------- */
/*      S = subprocessEnclosing(e)     */
          if((tM & /*fullEdgesMask(S)*/) | (tA & /*fullNodesMask(S)*/) != 0)
/* (2.1) IF S IS Internal */
            tM |= /*outgoingEdgesMask(S)*/;
/* (2.2) ELSE */
          	(tM, tA) = propagateEvent("Default", /*code(e)*/, /*index(e)*/, tM, tA);
/* (3) Case: Single upward --> Terminate, given S = subprocessEnclosing(e) ---------- */
 /* IF S IS Reusable FROM Non-Interrupting Boundary Event  */
          (tM, tA) = propagateEvent("Terminate", /*code(e)*/, /*index(e)*/, tM, tA);
 /* ELSE IF S IS Reusable */
          (tM, tA) = killProcess(0, tM, tA);
          (tM, tA) = propagateEvent("Default", /*code(e)*/, /*index(e)*/, tM, tA);
 /* ELSE */
          (tM, tA) = killProcess(/*index(S)*/, tM, tA);
          tM |= /*outgoingEdgesMask(S)*/;
/*  (4) Case: Broadcast --> Signal ------------------------------------------------------ */
          (tM, tA) = propagateEvent(/*type(e)*/, /*code(e)*/, /*index(e)*/, tM, tA); 
 		    }
     ...
 	\end{lstlisting}

The handling of error or escalation events follows the upward pattern, referred to as Case 1 in \autoref{lst:stepEvt}. At compilation time, we search for the corresponding catching event. This search is local to a reusable subprocess, i.e. only embedded subprocesses implemented in the contract of the reusable subprocess are traversed. If required, i.e. unsuccessful local search, the event is propagated to the parent contract by calling function {\tt propagateEvent} as shown in lines 7-9. 

The effect of this function depends on the type of event. For example, a \textit{terminate} event finishes the current process instance. An \textit{error} event may be caught by the process scope under which it occurs (which may lead to terminating the current instance), but it may also need to be propagated upwards to the parent process instance. Meanwhile, a signal event is broadcast down to any reachable children. In order to handle uncaught events, every process contract implements a function {\tt handleEvent}, which allows  a subprocess instance to indicate that an uncaught event has been generated under their scope.

Each process contract implements a function {\tt killProcess}, which terminates an instance of a subprocess given its index. In other words, it removes tokens on the arcs enclosed in the input subprocess and terminates any enabled child recursively. Generating the function {\tt killProcess} in a contract involves roughly the following steps. (i) Find the activities that, in addition to the root process (with index 0), can be interrupted in the contract, e.g., those with an interrupting boundary event attached. (ii)~For any possible activity to terminate, generate a block of instructions that validates if the index of the input subprocess matches to update the token marking accordingly. (iii) If the activity is reusable, then update {\tt startedActivities} and recursively terminate the related active instances, and later updatr the global field {\tt subInstanceStartedIndexes}. 

Continuing with the upward propagation, lines 10-20 in \autoref{lst:stepEvt} show how to handle an event caught internally. If the catching event is interrupting, two cases are distinguished: either the event starts an event-subprocess, or it is attached to the boundary of an activity. The former one terminates the current instance of the subprocess enclosing the event-subprocess, which is later restarted and enabled by adding a token in the outgoing arc of its start event. The second case terminates the subprocess where it is attached to the boundary event, which redirects the execution flow through its outgoing edge. Catching a non-interrupting event creates a new instance of the contract generated from that event in the same way described in \autoref{lst:step}, lines 42-46. In such a case, any related subprocess continues the execution unaltered. Besides, several catching events can be triggered as result of one propagation. The standard keeps open the behavior to follow if no subprocess catches a propagated event. Hence, Caterpillar interrupts any involved subprocesses in case of error propagation. Otherwise, the execution stays unchanged.

Cases 2 and 3 in \autoref{lst:stepEvt} illustrate the \textit{single upward} propagation. It occurs when a subprocess finishes by reaching a \textit{default end event} or a \textit{terminate} event. 
A \textit{default} event is thrown if no token exists in the {\tt marking} and no child subprocess is active, as checked in line 23. Here, {\tt fullEdgesMask} and {\tt fullNodesMask} represent the bitmasks of every edge/node in the subprocess containing the event $e$ to throw. Two cases are distinguished here: (2.1) $e$ is enclosed in an internal subprocess, i.e. embedded and (2.2) $e$ is enclosed in the reusable subprocess that defines the current a contract from which the event will be propagated to another contract. 
Case 3 handles the \textit{terminate} event, which terminates the subprocess instance in which it occurs and, in the case of a reusable sub-process, it propagates a ``Default'' event to its parent.

Case 4 in \autoref{lst:stepEvt} deals with signal events. This case involves calling function {\tt propagateEvent} repeatedly until reaching the root process. Once in the root, the event is disseminated to all subprocesses in the hierarchy by calling function {\tt broadcastSignal}. Every contract implements this function, which triggers any catching signal events and propagates the call to any reusable child recursively. Unlike upward propagation which comes from a child, catching/propagating a signal requires checking if the corresponding element is enabled. Note that boundary events and event-subprocesses do not fit in the typical token game because they have no incoming edges. Therefore, checking whether they are enabled relies on verifying the element to/in which they are attached/included depending on the case. Handling the signal when it is caught follows the same approach described above for error/escalation, depending on whether the event is interrupting or not.

To illustrate the approach, consider the example in \autoref{fig:nested-subprocesses}. Error event $e\_2$ enclosed in subprocess $S\_2$ is thrown in the {\tt step} function of the contract generated from $M$. This event is caught at the boundary of $S\_2$, which is embedded in $M$, and thus handled by the {\tt step} function which already started the propagation as outlined in Case 1.3 of \autoref{lst:stepEvt}. Consequently, the subprocesses $S\_2$ and $S\_3$ are terminated in the current instance of $M$, and the execution continues through the outgoing sequence flow of the boundary event with the same label ($e\_2$). On the other hand, the propagation of $e\_1$ starts in $M$ but is handled by the contract defined from $P$. To this end, the {\tt step} function calls {\tt propagateEvent} in $M$, which terminates the current instance and propagates the event by invoking function {\tt handleEvent} defined in $P$, which handles the event at the boundary of $M$. As a result, the remaining running instances of $M$ are terminated, and the execution continues via the outgoing flow of the boundary event.

 A smart contract generated from a non-root process always calls the function {\tt handleEvent} to propagate an event to its parent. This function checks the data received about the event (e.g., type, code, which child/instance is the source) to handle it accordingly. Indeed, the approach is similar to the one described in \autoref{lst:stepEvt}, i.e. trying to catch the event locally, if not possible then propagating it to the parent. However, a particular scenario must be solved when receiving a \textit{default} event from a contract encoding a multi-instance subprocess. Here, dealing with a sequential multi-instance involves checking if any reserved spot is empty (see \autoref{lst:stepProc}) to create a new instance. When no running instance exists, in both the parallel and sequential case, the execution continues by calling the function {\tt step} to add a token on the outgoing edge of the subprocess. For example, in \autoref{fig:nested-subprocesses}, if $P$ receives a default event from $M$, then the source instance is marked as finished. If there are no other running instances of $M$, then the function {\tt step} in $P$ is executed to add a token in the outgoing edge of $M$ (not present in \autoref{fig:nested-subprocesses}) and continuing the execution.
 
\section{Implementation and Evaluation}
\label{sec:impl-eval}

The Off-chain Runtime of Caterpillar and the web-based user interface are implemented in Node.js\footnote{https://nodejs.org/en/} and rely on the standard Solidity compiler solc-js\footnote{https://github.com/ethereum/solc-js} to compile the smart contracts. Interactions with running instances of the smart contracts are supported via the Ethereum client geth\footnote{https://github.com/ethereum/go-ethereum/wiki/geth}. The functionality of the Caterpillar's Off-chain Runtime is exposed via a REST API described in the following.  
Subsequently we describe the evaluation, which is focused on feasibility, correctness, and cost of our approach and implementation.
 
\subsection{REST API}

 Caterpillar's REST API is built around four types of resources summarized in \autoref{table:restAPI}: (i) \textit{models}, including BPMN models, the associated compilation artifacts and their instantiation, (ii)  \textit{processes}, which refers to process instances deployed, running or completed on the blockchain, and (iii- iv) \textit{worklists} and \textit{services} which involves interactions of user and external services with the running instances of a process. The component named ``Deployment Mediator'' in the architecture in \autoref{fig:architecture} exposes its functionalities through the resource \textit{models}, while the ``Execution Monitor'' responds to the REST actions involving the resources \textit{processes}, \textit{worklists} and \textit{services}. The component ``BPMN Compiler'' involves no REST interaction, but is invoked from the ``Deployment Mediator'' when a request to register a new model is received. Note that Caterpillar assumes the models created by the ``Modeling Panel'' are aimed to be deployed into the blockchain. Thus the ``Deployment Mediator'' triggers the compilation of such models before deploying them. 

\begin{table}[h!]
\centering
\begin{tabular}{|l|l|p{9cm}|}
\hline
\textbf{Verb} & \textbf{URI} & \textbf{Description} \\
\hline
POST & /models & Registers a BPMN model (triggers also code generation, compilation and updates the registry with a default configuration for the given model, i.e., after this operation the process model is ready to be instantiated, unless the operator incurred errors) \\
\hline
GET	& /models & Retrieves the list of registered BPMN models \\
\hline
GET	& /models/:m-hash & Retrieves a BPMN model and its compilation artifacts \\
\hline
POST & /models/:m-hash & Creates a new process instance from a given model \\
\hline
GET & /models/:m-hash/instances & Retrieves all the instances created from a given process model \\
\hline
GET & /processes/:p-address & Retrieves the current state of a process instance \\
\hline
PUT & /worklists/:wl-address/workitems/:wi-index & Checks-in a work item (i.e. user task) \\
\hline
PUT & /services/:s-address/tasks/:t-index & Executes a service task \\
\hline
\end{tabular}
\caption{Caterpillar's REST API.}
\label{table:restAPI}
\end{table}

 In the current version of Caterpillar, the addresses where the running smart 
contracts are linked to call activities and service tasks must be annotated 
in the corresponding process model. The smart contracts 
encoding worklists and factories are generated with a default policy which 
allows any external resource to instantiate a process and execute a started 
task. However, in future versions, the REST API and user interface will be 
extended to allow updates in the ``Runtime Registry'' to accordingly change 
the links/relations as well as to restrict the execution of a process to users 
with the corresponding privileges. In this direction, the ``Deployment 
Mediator'' will also serve as entry point to register and link process 
contracts, factories, worklists, and services, that are not necessarily 
produced by Caterpillar but relying on the structure outlined by its 
interfaces.

 A user can submit a BPMN model to be compiled and deployed using an HTTP 
POST request on the URL {\tt /models} which is also the approach followed by 
Caterpillar's ``Modeling Panel''. The request is made with a JSON message 
which includes the model serialized in the BPMN 2.0 XML standard format. 
Querying the process model metadata from the ``Distributed Repository'' uses 
HTTP GET requests, as shown in \autoref{table:restAPI}. An HTTP GET request 
on the URL {\tt /models} returns the information of all the models stored in 
Caterpillar's repository. Since this data can be voluminous, Caterpillar yields 
only a list containing the name and hash reference for each model stored in 
the repository. Then a user can retrieve all the information associated with 
a particular model (i.e. serialized BPMN model, solidity code, etc.) using an 
HTTP GET with the model's corresponding hash.
 
 Given the identifier of a process, i.e. hash produced when stored in the 
repository, a user can use an HTTP POST to request the creation of an 
instance. After submitting the transactions to create and start the instance,
Caterpillar retrieves the URL where the status of the newly created instance 
can be accessed. In the 
current implementation, the URL associated with a process instance includes 
the address of the underlying contract. Moreover, once the corresponding 
transactions are included in the blockchain, the ``Runtime Registry'' publishes a 
solidity event with the address of the new instance in the Ethereum Log. 
Thus, the ``Event Monitor'' accordingly sends notifications with this address 
to the ``Execution Monitor'' and ``Execution Panel''.  
 
 At any time, a user can query the state of a process instance by using an 
HTTP GET request on the URL associated with the instance of interest. In 
response, a JSON message is sent, with the information required to visualize 
and execute any started user/service task. By way of example, consider the ``
Order to cash'' process model presented in \autoref{fig:running} and assume 
the execution of a process instance has progressed up to the point where 
two instances of the subprocess ``Carrier Selection'' have started (\autoref{fig:running}(b)). Moreover, 
consider that a participant executed the task ``Request quote'' in one 
instance of the subprocess ``Carrier selection''. \autoref{lst:state} shows 
the JSON message sent responding to the HTTP GET request on the URL {\tt 
/processes/o2c-address}, where the acronym ``o2c'' references in a compact 
way the root process ``Order to Cash''. 
 
 
 \begin{lstlisting}[float, language=json,label=lst:state,caption=Sample process instance's state (model from \autoref{fig:running}).]
 {
		"process-identifier": "o2c-hash",
		"href": "/processes/o2c-address",
		"workitems": [
			{
				"elementId": "Request_Quote_Id",
				"name": "Request_Quote",
				"importParameters": [
					{ "type": "uint", "name": "quote" }
				],
				"instances": [
						{
							"exportParameters": [],
							"href": "/worklists/wl_address/workitems/wi_1",
						}
				]
			}, 
			{
				"elementId": "Submit_Quote_Id",
				"name": "Submit_Quote",
				"importParameters": [],
				"instances": [
					{
						"exportParameters": ["type": "uint", "name": "quote", "value": "100"],
						"href": "/worklists/wl_address/workitems/wi_3"
					}
				]
			}
		]
		"services": []
 }
 \end{lstlisting}

First, the JSON message in \autoref{lst:state} provides the hash identifier 
and the URL to access the process instance. Next, two lists named workitems 
and services contain the information of started user and service tasks 
respectively. For each element, its identifier and name in the BPMN model are 
retrieved. For the workitems, the import parameters are listed, which must be provided by a 
participant when executing the task. Note that the type and name of such 
parameters are used by the ``Execution Panel'' to generate web forms that the 
user must fill. Finally, there is a list with the URLs to execute the task through an 
HTTP PUT and the parameters to export, whose values must be displayed when 
visualizing the task. Note that the export parameters are derived from information 
stored in the smart contract, which varies among instances. For example, the 
quote provided by a shipper when executing the task ``Submit quote'' -- which 
also starts the task ``Request quote'' -- is sent to the carrier who can decide 
whether to submit such a quote or not. The addresses where the instances of the 
subprocesses ``Goods Shipment'' and ``Carrier Selection'' are running are not 
needed (and not returned), because Caterpillar forces user tasks to be executed through the 
corresponding worklist.
  
As indicated above, the execution of a user task requires an HTTP PUT to a 
URL provided to that end when querying the process state, e.g. {\tt /worklists/wl\_address/workitems/wi-\_1} in line 14 of \autoref{lst:state}. Such URL 
contains as parameters the address when running the worklist and the index of 
the corresponding workitem. Furthermore, the request needs to include a JSON message with the expected values for the task. 
 

\subsection{Evaluation} 

%

This section presents an experimental evaluation aimed at assessing the cost of executing business processes using Caterpillar, relative to other baselines that either only record the execution of the process (without enforcing it) or that do not fulfill the design principles outlined in Section~\ref{ssec:designphilosophy}.
In the following, we describe the datasets, experimental setup, and findings.



\subsubsection{Datasets}

In line with our earlier work on blockchain-based collaborative process execution~\cite{GarciaPDW17}, we used four datasets for the evaluation, each of them consisting of a BPMN process model and corresponding event log.  
\autoref{tbl:datasets} presents the statistics of the datasets.
The first dataset, referred to as \textit{Invoicing},  is an event log of a real-world business process, used and distributed by Minit\footnote{http://www.minitlabs.com/ - last accessed 17/05/2018} for demonstrating its process mining tool. The BPMN model for this dataset was directly derived from the
event log, using a state-of-the-art process discovery tool~\cite{AugustoCDRB16}. 
Since the BPMN model is automatically discovered, some traces ($<1\%$) were non-conforming traces. We discarded these non-conforming traces, since they represent a marginal subset of all traces.

The other three datasets, i.e. \textit{supply chain}, \textit{incident management} and \textit{insurance claim},
are process models extracted from the literature and were used in the experiments reported in~\cite{WeberXRGPM16}.
In this case, the event logs for the experiments were generated after the BPMN models using a simulation tool.
In order to assess Caterpillar's ability to differentiate between confirming and non-conforming inputs, we inserted noise into these event logs with non-conforming traces, by randomly changing events over some of the traces also selected at random.  


\begin{table}[h!]
\center
\begin{tabular}{|p{5.0em}|c|c|l|c|}
\hline
\textbf{Process} & \textbf{Tasks} & \textbf{Gateways} & \textbf{Trace Type} & \textbf{Traces} \\
\hline
Invoicing & 40 & 18 & Conforming & 5,317 \\
\hline
\multirow{2}{4.2em}{Supply chain} & \multirow{2}{*}{10} & \multirow{2}{*}{2} & Conforming & 5 \\
\cline{4-5} 
&&& Not conforming & 57 \\
\hline
\multirow{2}{4.2em}{Incident Management} & \multirow{2}{*}{9} & \multirow{2}{*}{6}
& Conforming & 4 \\
\cline{4-5}
&&& Not conforming & 120 \\
\hline
\multirow{2}{4.2em}{Insurance claim} & \multirow{2}{*}{13} & \multirow{2}{*}{8}
& Conforming & 17 \\
\cline{4-5}
&&& Not conforming & 262 \\
\hline
\end{tabular}
\vspace{-0.3em}
\caption{Datasets used in the evaluation.}
\label{tbl:datasets}
\vspace*{-6mm}
\end{table}

\subsubsection{Experimental setup}


We replayed the distinct log 
traces and interacting with Caterpillar using its REST API. To that end,
we implemented a replayer component, which compiles, configures and
deploys each one of the four process models in the input datasets. After compiling and deploying the process models, the replayer reads
the corresponding event log and sequentially feeds each the events 
in the log into the Caterpillar system through its REST API.
To execute a task, the replayer queries the state of the process instance,
formats a JSON message with the required data, and submits the request to
Caterpillar.
Then, the replayer waits for a notification from Caterpillar's ``Event Monitor'',
indicating that the block containing the underlying transaction has been completed,
and additional information that includes the transaction hash and gas consumption.
The replayer queries Caterpillar, which checks the on-chain runtime to determine if there is a work-item matching the next event in the event log. If the response to the replayer indicates that there is no such work-item (meaning that the on-chain runtime rejects the event in question), the replayer marks the trace as non-conforming, skips the event, and continues reading the next event in the log. 


For comparison, we run the same experiments using three baselines. 
The first baseline, namely \emph{Basic} is designed to reflect the approach where the process is executed in one or more off-chain BPMSs, and the blockchain is only used to leave tamper-proof execution  traces of the process via a connector between the BPMS and a blockchain platform~\cite{bonitasoft,IBMBlockchainBPM}. 
In this baseline, the smart contract of a process instance records a reference to each event executed in this process instance, using a dynamic array of bytes32. Here, we assume the data generated by each event is stored off-chain.
The second baseline, namely \emph{Default} corresponds to a smart contract that enforces the control flow of the process and stores the data required to evaluate the conditions in the decision gateways of the BPMN model, but without any work-item management (i.e.\ no handling of user and service tasks) and with all runtime components left off-chain. In other words, this approach does not satisfy design principle \# 4 in Section~\ref{ssec:designphilosophy} -- the smart contracts generated from the process model rely on other off-chain runtime components. This approach is used proposed in~\cite{GarciaPDW17}.
The third baseline, namely \emph{Optimized}, is the same as \emph{Default} with some code optimizations to reduce the number of bits required to store the state of process instances, as outlined in~\cite{GarciaPDW17}.
%

The experiments were run on a personal computer with an Intel i5-5200 dual-core CPU.
Moreover, we run them over \textit{testrpc}\footnote{https://github.com/0xProject/testrpc},
which is a NodeJS-based implementation of Ethereum client used for development purposes.
In this way, both Caterpillar's runtime and log replayer ran over the same computer.


\subsubsection{Results and discussion}

Given that gas consumption is deterministic, the traces were grouped together such that only 
one execution was performed for each distinct trace in the event log. The same approach 
was taken in our previous work
\cite{GarciaPDW17}.
As expected, all non-conforming behavior was properly handled by Caterpillar:
the request corresponding to a non-conforming event was ignored by Caterpillar
because the underlying task was not enabled (lifecycle state: ``executing'').


The measurements of gas consumption are shown in~\autoref{tbl:results}. The table reports the cost of instantiation of the process (in gas) and the runtime cost, which means the cost of handling all the events related to a given process instance.
The costs reported in this table are adjusted such that they reflect the average overall gas consumption. 
The last two columns of the table report the relative overhead of Caterpillar relative to each of the baselines. For example, an overhead of 3.51 for the instantiation cost implies that Caterpillar consumes 3.51 times more gas that the baseline in question.
        
\begin{table}[h!]
\center
\newcommand{\mywidth}{6.3em}
\begin{tabular}{|p{6.0em}|c|l|c|c|c|c|}
\hline
{\bf Process} & \parbox[t][10pt][t]{1.10cm}{\bf Tested Traces} &  \parbox[t][10pt][t]{1.20cm}{\bf Translator Version} & \multicolumn{2}{c|}{\bf W. Avg. Cost}
& \multicolumn{2}{c|}{\bf Relative Overhead} \\
\cline{4-7}
&&&{\bf Instant.} & {\bf Exec.} & {\bf Instant.} & {\bf Exec.} \\
\hline
\multirow{4}{\mywidth}{Invoicing} & \multirow{4}{*}{5316} & Basic 
& 123,625 &  589,228
& 22.9 & 1.8 \\
\cline{3-7}
&& Default 
&  1,089,000 & 383,109 
& 2.60 & 2.84 \\
\cline{3-7}
&& Optimized 
&  807,123 & 297,351 
& 3.51 & 3.66 \\
\cline{3-7}
&& \bf Caterpillar 
& \bf 2,830,063 & \bf 1,088,315
& -- & -- \\
\hline
\multirow{4}{\mywidth}{Supply chain} & \multirow{4}{*}{62} & Basic 
& 123,625 & 570,444
& 8.9 & 0.99 \\
\cline{3-7}
&& Default 
& 304,084 & 281,206 
& 3.62 &  2.02 \\
\cline{3-7}
&& Optimized 
& 298,564 & 272,186 
&  3.69  &  2.08 \\
\cline{3-7}
&& \bf Caterpillar 
&  \bf 1,100,590 & \bf 566,861
& -- & --\\
\hline
\multirow{4}{\mywidth}{Incident mgmt.} & \multirow{4}{*}{124} 
&  Basic 
& 123,625 & 375,929
& 9.0 & 0.86 \\
\cline{3-7}
&& Default 
&  365,207 & 185,680 
&  3.07 &  1.75 \\
\cline{3-7}
&& Optimized 
&  345,743 & 166,345 
&  3.24 &  1.95 \\
\cline{3-7}
&& \bf Caterpillar 
& \bf 1,119,803 & \bf 324,420
& -- & -- \\
\hline
\multirow{4}{\mywidth}{Insurance claim} & \multirow{4}{*}{279} 
& Basic 
& 123,625 & 1,008,840
& 10.8 & 1.23 \\
\cline{3-7}
&& Default 
& 439,143 & 552,274 
&  3.05 &  2.24 \\
\cline{3-7}
&& Optimized 
& 391,510 & 514,712 
& 3.42 &  2.40 \\
\cline{3-7}
&& \bf Caterpillar 
& \bf 1,338,152 & \bf 1,235,617
& -- & -- \\
\hline
\end{tabular}
\vspace{-0.3em}
\caption{Experiment results.}
\label{tbl:results}
\vspace*{-6mm}
\end{table}

As expected, \autoref{tbl:results} shows that the cost of instantiation of the Basic approach is considerably lower than all other approaches. This is because the smart contract generated by this approach is relatively small: it the contract exposes a function that takes an event as input and simply records it on-chain.
On the other hand, the execution costs of this contract is comparable to that of Caterpillar since the Basic contract has to make a write operation on a dynamic array, for each event. In contrast, the Default and Optimized approaches do not store each event, but instead they store the state of the process instance in a bit-set, which is a less costly operation. Caterpillar does the same, but it also performs other operations in addition to updating the state of the process instance. 

\autoref{tbl:results}  also shows that, on average, the smart contracts generated by Caterpillar consume two to three times
more gas than that required by the solidity code generated by \emph{Default}
and \emph{Optimized}. This trend was expected, as the code generated by
\emph{Default} and \emph{Optimized} is encapsulated in a single smart contract
and deals only with basic control flow and data flow perspectives. In contrast,
due to more advanced architectural design, Caterpillar produces several smart
contracts to support concerns such as workitem handling and runtime housekeeping
among other things. One single interaction between Caterpillar's 
off-chain and on-chain components may involve chains of interactions between
up to four solidity smart contracts, e.g. 
\textit{workflow}, \textit{worklist}, \textit{factory} and \textit{registry}.
Therefore, it is worth noting that the comparison between the costs reported
for Caterpillar and those reported for \emph{Default} and \emph{Optimized} is
not straightforward, because the functionality provided by Caterpillar is
more sophisticated than that provided by the other two options.
Moreover, the costs associated to Caterpillar's code can still be reduced
by applying the techniques described in~\cite{GarciaPDW17}, which are those
that are associated with \emph{Optimized}. The latter is left as a venue for
future research.




\subsection{Source code}

The source code of Caterpillar can be downloaded under the BSD 3-clause 
``New'' or ``Revised'' License from \url{https://github.com/orlenyslp/Caterpillar}.
Note that the repository includes two versions of Caterpillar and that version 2.0
is the one implementing the architecture and components described in this article.
~\footnote{Neither the publication of this paper nor the software release 
should be construed as granting any rights in relation to patents or patent 
applications held by CSIRO or by the authors of this paper.} 

Caterpillar's code distribution contains two folders. The first named 
\emph{caterpillar core} includes the implementation of the ``Off-chain Components'' (
``BPMN Compiler'', ``Deployment Mediator'', ``Execution Monitor'' and ``Event 
Monitor''). Moreover, the sub-folder ``Abstract'' includes the Solidity 
interfaces to be implemented in each contract generated from a BPMN model. 
This folder also includes the full implementation of the ``Runtime 
Registry''. The second folder named \emph{execution panel} implements the web 
application serving as the user interface. The repository contains the
instructions to get all dependencies (i.e. third party libraries) and to run 
Caterpillar. For convenience, a Docker image has been built
that provides a pre-configured version of Caterpillar ready to be used.
The instructions for installing the Docker image are also provided in the
code repository. Finally, the repository also contains a number of
examples (i.e. BPMN models) that showcase the coverage of BPMN constructs
supported by Caterpillar, are ready for deployment to Caterpillar and
serve as demonstration of its functioning.


\section{Conclusion}
\label{sec:outlook}

This article presented the design and implementation of the Caterpillar system for blockchain-based execution of collaborative business processes captured in the  BPMN notation. 
To the best of our knowledge, Caterpillar is the first blockchain-based process execution engine capable of handling process models with subprocesses, as well as advanced BPMN constructs such as boundary events and multi-instance activities.
Together with our previous proposal~\cite{GarciaPDW17}, of which it is a major extension, Caterpillar is the first system where the entire collaborative process is executed on the blockchain in the sense that all the state of the process instances and their links are maintained ``on-chain'', and all the control-flow logic is encoded in smart contracts. 
Similarly, Caterpillar is the first system that does not assume that the parties in the process use message exchanges for coordination, but instead, the parties use the blockchain as the only coordination mechanism.
In this respect, the Caterpillar system advances the understanding of how to combine the high-level abstractions of existing (intra-organizational) BPMSs with the trust-enhancing capabilities of blockchain technology in order to support the execution of collaborative business processes.

The presented version of Caterpillar provides a generic worklist handler, 
which does not implement any access control mechanism, meaning that any party 
can alter the state of execution of any process instance (e.g. they can mark 
a task as ``completed''). Access control is a core functionality of any BPMS, 
and is particularly important in collaborative business processes involving 
untrusted parties. Accordingly, our main avenue for future work is to extend 
Caterpillar with a worklist handler that implements a suitable access control 
mechanism for collaborative processes. 
Mainstream BPMSs are based on a Role-Based Access Control (RBAC) model, where 
each task in the process model is mapped to a role and any user (e.g.\ 
process worker) who plays the role corresponding to a task can perform any 
instance of this task. Some BPMSs support additional access constraints such 
as ``retain familiar'' or the ``four-eyes principle'', as well as delegation~
\cite{RussellAHE05}. 
Collaborative business processes, however, require more sophisticated access 
control mechanisms. In particular, some collaborative processes in the field 
of logistics require dynamic binding and re-binding. For example, in a buyer-
supplier-carrier process, the carrier might sometimes be appointed by the 
supplier, but other times by the buyer. Moreover, the responsibilities of the 
actors in a collaborative business process are in some cases determined only 
at runtime, when a case is already running. For example, sometimes the buyer 
is responsible for paying the shipping costs, while at other times these 
costs are borne by the supplier. In some cases, the seller may have the right 
to change the carrier after its initial appointment, for example if 
the initially appointed carrier is not able to pick up the merchandise on 
time, or if the products to be shipped are not yet ready and the shipment 
needs to be postponed. Moreover, carriers might sub-contract to other 
carriers or transportation providers. How to best support these scenarios is 
an open question.


While the experimental evaluation show that Caterpillar can handle realistic process models, it also suggests that the approach would not scale to very large process models with hundreds or thousands of elements, particularly when running on a public blockchain. Higher scalability could be achieved by using consortium blockchain technology, such as Hyperledger or Ethereum Proof-of-Authority consensus, which support much higher throughput. Investigating the performance of the Caterpillar approach on different blockchain platforms and configurations is a direction for future work.

 \section*{Acknowledgment}
This work is partly funded by the Estonian Research Council (grant IUT20-55).

\bibliography{references}

\begin{thebibliography}{10}

\bibitem{WeberXRGPM16}
Weber I, Xu~X, Riveret R, Governatori G, Ponomarev A, Mendling J. {Untrusted
  Business Process Monitoring and Execution Using Blockchain}.  In: Proceedings
  of 14th International Conference on Business Process Management, {BPM} 2016,
  Rio de Janeiro, Brazil; 2016.

\bibitem{DBLP:journals/tmis/MendlingWABCDDC18}
Mendling J, Weber I, Aalst W.M.P, et al. Blockchains for Business Process
  Management - Challenges and Opportunities.  {\it {ACM} Trans. Management Inf.
  Syst.. }2018;9(1):4:1--4:16.

\bibitem{UKBlockchainReport}
{UK Government Chief Scientific Adviser} . {\it Distributed Ledger Technology:
  Beyond Block Chain. }: UK Government Office of Science; 2016.

\bibitem{aureport}
Staples Mark, Chen Shiping, Falamaki Sara, et al. {\it Risks and opportunities
  for systems using blockchain and smart contracts. }: Data61(CSIRO)Sydney;
  2017.

\bibitem{Gartner:2018:CIOSurveyBC}
Release Gartner~Press. Gartner Survey Reveals the Scarcity of Current
  Blockchain Deployments  \url{https://www.gartner.com/newsroom/id/3873790} --
  last accessed 22-Jun-2018; .

\bibitem{bpmnspec}
{Object Management Group} . {BPMN 2.0 Specification},
  \url{http://www.omg.org/spec/BPMN/2.0/} -- last accessed 22-Jun-2018; .

\bibitem{GarciaPDW17}
Garc{\'{\i}}a{-}Ba{\~{n}}uelos L, Ponomarev A, Dumas M, Weber I. Optimized
  Execution of Business Processes on Blockchain.  In: Proceedings of 15th
  International Conference on Business Process Management, {BPM} 2017,
  Barcelona, Spain; 2017.

\bibitem{Prybila:FGCS:2017}
Prybila C, Schulte S, Hochreiner C, Weber I. Runtime Verification for Business
  Processes Utilizing the {Bitcoin} Blockchain.  {\it Future Generation
  Computer Systems (FGCS). }2017;.

\bibitem{Lopez:BPM-Demo:2017}
L{\'o}pez-Pintado O, Garc{\'i}a{-}Ba{\~{n}}uelos L, Dumas M, Weber I.
  {Caterpillar: A Blockchain-Based Business Process Management System}.  In:
  Proceedings of the {BPM} Demo Track and {BPM} Dissertation Award co-located
  with 15th International Conference on Business Process Modeling {(BPM} 2017),
  Barcelona, Spain; 2017.

\bibitem{2019-Blockchain-Book}
Xu~Xiwei, Weber Ingo, Staples Mark. {\it Architecture for Blockchain
  Applications}.
\newblock Springer; 2019.

\bibitem{wireprotocol}
{The Ethereum Foundation} . {Ethereum Wire Protocol},
  \url{https://github.com/ethereum/wiki/wiki/Ethereum-Wire-Protocol} -- last
  accessed 22-Jun-2018; .

\bibitem{jsonrpc}
{JSON-RPC Working Group} . {JSON-RPC 2.0 Specification},
  \url{http://www.jsonrpc.org/specificationl} -- last accessed 22-Jun-2018; .

\bibitem{omg2013bpmn}
OMG . {\it {Business Process Model and Notation (BPMN), Version 2.0.2}. } 2013.

\bibitem{Hull:ICSOC:2016}
Hull R, Batra V.S, Deutsch Y~Chenband~A, III F.F.T~Heath, Vianu V. Towards a
  Shared Ledger Business Collaboration Language Based on Data-Aware Processes.
  In: Proceedings of 14th International Conference on Service-Oriented
  Computing, {ICSOC} 2016, Banff, AB, Canada; 2016.

\bibitem{Nigam:IBM-SJ:2003}
Nigam A, Caswell N.S.. Business Artifacts: An Approach to Operational
  Specification.  {\it IBM Syst. J.. }2003;42(3):428--445.

\bibitem{Norta2015}
Norta A. Creation of Smart-Contracting Collaborations for Decentralized
  Autonomous Organizations.  In: Proceedings of 14th International Conference
  on Perspectives in Business Informatics Research, {BIR} 2015, Tartu, Estonia;
  2015.

\bibitem{Frantz:ECAS:2016}
Frantz C, Nowostawski M. From Institutions to Code: Towards Automated
  Generation of Smart Contracts.  In: 1st International Workshops on
  Foundations and Applications of Self* Systems (FAS*W), Augsburg, Germany;
  2016.

\bibitem{sturm2018lean}
Sturm Christian, Szalanczi Jonas, Sch{\"o}nig Stefan, Jablonski Stefan. A Lean
  Architecture for Blockchain Based Decentralized Process Execution.  In: ;
  2018.

\bibitem{2018-Tran-BPM-Demo}
Tran An~Binh, Lu~Qinghua, Weber Ingo. Lorikeet: A Model-Driven Engineering Tool
  for Blockchain-Based Business Process Execution and Asset Management.  In:
  BPM'18: International Conference on Business Process Management, Demo track;
  2018; Sydney, NSW, Australia.

\bibitem{Rikken:bpmleader.com:2015}
Rikken O. {\it {BPM} and Blockchain, miles apart or closer than you think?. }
  \url{https://www.bpmleader.com/2015/11/17/bpm-blockchain-miles-apart-closer-think/}
  -- last accessed 22-Jun-2018; 2015.

\bibitem{bonitasoft}
Palacin L. {Accelerate blockchain technology adoption with Bonita BPM and Chain
  Core,}  \url{https://vimeo.com/202058656} -- last accessed 22-Jun-2018; .

\bibitem{IBMBlockchainBPM}
Auberger Larissa, Kloppmann Matthias. {\it Digital process automation with
  {BPM} and blockchain, Part 1: Combine business process management and
  blockchain. } 2017.

\bibitem{SadiqGN07}
Sadiq Shazia~Wasim, Governatori Guido, Namiri Kioumars. Modeling Control
  Objectives for Business Process Compliance.  In:  Lecture Notes in Computer
  Science, vol. 4714: :149--164Springer; 2007.

\bibitem{ethereumoracles}
Buterin V. {Ethereum and Oracles},
  \url{https://blog.ethereum.org/2014/07/22/ethereum-and-oracles/} -- last
  accessed 22-Jun-2018; .

\bibitem{AugustoCDRB16}
Augusto A, Conforti R, Dumas M, Rosa M~La, Bruno G. Automated Discovery of
  Structured Process Models: Discover Structured vs. Discover and Structure.
  In: Proceedings of 35th International Conference on Conceptual Modeling, {ER}
  2016, Gifu, Japan; 2016.

\bibitem{RussellAHE05}
Russell N, Aalst W.M.P., Hofstede A.H.M., Edmond D. Workflow Resource Patterns:
  Identification, Representation and Tool Support.  In: Proceedings of of
  CAiSE; 2005.

\end{thebibliography}

\end{document}